\documentclass[acmsmall, authorversion]{acmart}

\usepackage[utf8]{inputenc}
\usepackage{acronym}
\usepackage{graphicx}
\usepackage{setspace}
\usepackage{textcomp}
\usepackage{ragged2e}
\usepackage{amsmath}
\usepackage{tabularx}
\usepackage{float}
\usepackage{verbatim}
\usepackage{array}
\usepackage{color}
\usepackage{subfigure}
\usepackage{balance}
\usepackage{hyperref}
\usepackage{pifont}

\newcommand{\sol}{\emph{MAGNETO}}
\newcolumntype{P}[1]{>{\centering\arraybackslash}p{#1}}

\newcommand{\cmark}{\ding{51}}%
\newcommand{\xmark}{\ding{55}}%

\acrodef{HID}{Human Interface Device}
\acrodef{RF}{Radio Frequency}
\acrodef{RF-DNA}{Radio Frequency-Distinct Native Attribute}
\acrodef{AWGN}{Additive White Gaussian Noise}
\acrodef{MDA/ML}{Multiple Discriminant Analysis-Maximum Likelihood}
\acrodef{USB}{Universal Serial Bus}
\acrodef{DNS}{Domain Name System}
\acrodef{OFDM}{Orthogonal Frequency-Division Multiplexing}
\acrodef{URE}{Unintentional Radiated Emissions}
\acrodef{IRE}{Intentional Radiated Emissions}
\acrodef{RSS}{Received Signal Strength}
\acrodef{FFT}{Fast Fourier Transform}
\acrodef{SDR}{Software Defined Radio}
\acrodef{ROC}{Receiver Operating Characteristic}
\acrodef{PCB}{Printed Circuit Board}
\acrodef{OS}{Operating System}
\acrodef{SVM}{Support Vector Machine}
\acrodef{TPR}{True Positives Ratio}
\acrodef{FPR}{False Positives Ratio}

\setcopyright{acmcopyright}
\copyrightyear{2020}
\acmYear{2020}
\acmDOI{10.1145/1122445.1122456}

\acmJournal{TECS}

\begin{document}

\title{MAGNETO: Fingerprinting USB Flash Drives via Unintentional Magnetic Emissions}

\author{Omar Adel Ibrahim}
\email{oaibrahim@hbku.edu.qa}

\author{Savio Sciancalepore}
\email{ssciancalepore@hbku.edu.qa}

\author{Gabriele Oligeri}
\email{goligeri@hbku.edu.qa}

\author{Roberto Di Pietro}
\email{rdipietro@hbku.edu.qa}


\begin{abstract}
Universal Serial Bus (USB) Flash Drives are nowadays one of the most convenient and diffused means to transfer files, especially when no Internet connection is available.
However, USB flash drives are also one of the most common attack vectors used to gain unauthorized access to host devices. For instance, it is possible to replace a USB drive so that when the USB key is connected, it would install passwords stealing tools, root-kit software, and other disrupting malware. In such a way, an attacker can steal sensitive information via the USB-connected devices, as well as inject any kind of malicious software into the host.

To thwart the above-cited raising threats, we propose \mbox{MAGNETO}, an efficient, non-interactive, and privacy-preserving framework to verify the authenticity of a USB flash drive, rooted in the analysis of its unintentional magnetic emissions. We show that the magnetic emissions radiated during boot operations on a specific host are unique for each device, and sufficient to uniquely fingerprint both the brand and the model of the USB flash drive, or the specific USB device, depending on the used equipment. Our investigation on 59 different USB flash drives---belonging to 17 brands, including the top brands purchased on Amazon in mid-2019---, reveals a minimum classification accuracy of $98.2$\% in the identification of both brand and model, accompanied by a negligible time and computational overhead. \mbox{MAGNETO} can also identify the specific USB Flash drive, with a minimum classification accuracy of $91.2$\%.
Overall, \mbox{MAGNETO} proves that unintentional magnetic emissions can be considered as a viable and reliable means to fingerprint read-only USB flash drives. Finally, future research directions in this domain are also discussed.
\end{abstract}




\maketitle

\section{Introduction}
\label{sec:intro}

The \ac{USB} standard is nowadays the most convenient, cheap, and widespread means to physically connect peripheral devices to workstations and laptops \cite{usb3dot0}. Over the last years, the \ac{USB} standard replaced a large number of earlier interfaces, including serial and parallel ports, used to supply either power or data connectivity to external devices, thus providing a unique and standardized interface to be used for the connection of any peripheral board. 
In addition, the ongoing introduction of the USB 3.0 standard specification is further widening the application scenarios where USB connections are suitable, thanks to transfer speeds up to 5 Gbps and a maximum storage capacity of 1 TB \cite{Lyu2019}. As a result, USB connectivity is more and more used by companies in industrial environments to transfer data between devices, as well as to charge electrical components \cite{Prz2016}.

The widespread diffusion of USB connections is confirmed by the increasing market size associated with the new wave of USB 3.0 devices. According to Transparency Market Research, the worldwide market for USB 3.0 drives is growing at a Compound Annual Growth Rate (CAGR) of roughly 23.5\% over the last years, reaching 3.1 billion US\$ in 2020, from 1.1 billion US\$ in 2015. At the same time, USB devices are getting cheaper: the average price of a USB drive in China decreased from 6.7 USD/Unit in 2011 to 6.0 USD/Unit in 2016, and a further decreasing trend is expected in the upcoming years \cite{tmr2017}.   

Despite the provided advantages, the USB standard communication interface also poses increasing concerns from the security perspective \cite{Noyes2016}. Being focused on boosting the availability of connected peripherals and on reducing the latency in accessing remotely stored data, the USB software design paid little attention to security issues. Exploiting such a weak security barrier, many attacks and vulnerabilities emerged over the last few years \cite{Nissim2017}. 
Indeed, malicious attackers have been able to modify the firmware of specific (faulty) USB flash drives, turning them into a vector to launch tools for information stealing, host system impairment, Internet traffic redirection, and malware injection, to name a few \cite{BadUSB}, \cite{Mulliner2012}. As reported by the latest media and newspapers, USB attacks are very likely to happen especially when the target system is not connected to the Internet, such as in the case of ships, remote keyless entry systems, and critical infrastructures~\cite{ships2018},~\cite{ss2019}.

While a few countermeasures have been provided in the literature---mainly based on standard anti-malware software---attackers are still able to escape detection, e.g., by masquerading the malicious USB device as a legitimate one at the time of connection to the host \cite{Tian2016}. Even if the forthcoming USB Type-C standard enhanced the security guarantees via digital certificates, installing and verifying such certificates into regular USB 2.0 and 3.0 devices would lead to a complete software and hardware redesign of the interfaces, thus being invasive and hard to be deployed. Especially in industrial companies and critical infrastructure scenarios, the impact of such attacks can be devastating, preventing the daily activities of the company and possibly leading to huge economic losses on the whole country relying on the target critical infrastructure.

To provide an effective tool for the detection of malicious USB devices, in this paper we present \sol, a framework able to identify replaced and faulty USB flash drives connected to a specific host, based on unintentional magnetic emissions radiated by the USB boards, during the execution of the boot procedure. 
We show that the magnetic emissions radiated at very low frequencies by USB flash drives immediately after the connection to a specific host system are consistent and unique, i.e., they show the same profile over time, and they are specific to the model and version of the analyzed device. 
Thus, such low-frequencies magnetic emissions can be effectively used to achieve both a method for USB brand and model identification and a physical-layer fingerprinting mechanism for USB devices---hence providing a powerful tool to spot the replacement or malicious corruption of the authorized USB flash drive.

Our experimental campaign took into consideration 59 different USB devices, including the top brands purchased on Amazon in July 2019, and it revealed impressive performances. When deployed using low-cost \acp{SDR}, \sol\ reaches up to $98.2$\% of accuracy in correctly identifying the brand and model of the device, after an observation time of only 1.08 seconds. When the identification of the single USB flash drive is crucial, such as in critical infrastructures, using a wide-band spectrum analyzer \sol\ can discriminate the specific USB device connected to a given host with a minimum accuracy of the $91.2$\%. Such outstanding performances require only minor computations on a commercial laptop, further confirming the viability of \sol.

It is worth noting that \sol\ could be extended also beyond the execution of the boot procedure. In fact, by just extending the time duration of the observation window, it could be possible to monitor the activities of the USB flash drive when it is in the idle state, thus detecting any asynchronous (potentially malicious) activities originated by the USB device.

\sol, though being of general applicability, could benefit those environments interested in strictly protecting the access to sensitive equipment---such as critical infrastructures and industrial companies. 
Indeed, under the reasonable assumption that only a few (brands of) USB flash drives are authorized (e.g., the ones distributed by the company itself), \sol\ can be implemented on a reference computing machine, to detect any replacement or modification of the intended USB flash drives, further strengthening the enforcement of security policies.

We highlight that three main elements distinguish \sol\ from previous contributions. First, \sol\ is the first solution to exploit the usage of unintentional \emph{magnetic emissions} to address the threats associated with malicious USB drives. Despite this feature could appear just an application of a well-known background notion, it brings new unprecedented challenges. Indeed, differently from other devices analyzed in the scientific literature, USB flash drives cannot be modified (or deeply inspected) by end-users, and they should be analyzed as \emph{black-box}.

Second, we prove that unintentional magnetic emissions occurring during the boot process of a USB flash drive can be used to fingerprint its hardware, without any intervention at either the firmware or the software level, but only requiring to plug-in the USB flash drive. 
Further, our solution guarantees the full privacy of the device under test, making \sol\ overall enjoying the properties of being non-invasive, minimal-interactive, and privacy-preserving, and fully-compatible with read-only devices. To the best of our knowledge, all these features are not provided by any other competing solution in the literature.

Finally, the open-source code and data of \sol\ framework are available at \cite{crilab}.
This will allow practitioners, industries, and academia to verify our claims, as well as to extend the adoption of \sol\ as a ready-to-use basis for their further software development.

The sequel of this paper is organized as follows: Section \ref{sec:background_related} provides a brief background on unintentional RF and magnetic emissions, as well as an overview of USB security-related work; Section \ref{sec:scenario_adv_model} highlights the assumed scenarios and the adversary model; Section \ref{sec:idea} provides the details of \sol; Section \ref{sec:experiments} describes the tools and the results achieved by using \sol; Section \ref{sec:discussion} highlights the advantages and limitations of \sol; and, finally, Section \ref{sec:conclusion} tightens conclusions.

\section{Background and Related Work}
\label{sec:background_related}

Despite being assembled in a \ac{PCB} through industrial-grade processes, any embedded device produces even small unintentional magnetic emissions, emanated directly from the integrated circuits. 

The source of such emissions is mainly due to the local oscillator, that continuously emits electromagnetic waves at a given fundamental frequency, to provide a unique timing reference for hardware and software operations executed by the embedded device. Even being narrow-band, the signal emitted by the oscillator creates also several \emph{harmonics}, i.e., components at frequencies multiple than the fundamental tone, characterized by a non-negligible power spectral density. Such harmonics couple with very small integrated wires on the embedded device and create resonating effects, that turn the wires into small antennas, emanating RF power \cite{Acharya2013}. The physical effects underpinning these phenomena are described by the physics fundamental Maxwell equations, and also specific mathematical models are available in the literature, describing their complex creation \cite{bole2009}. On the one hand, this phenomenon can be the cause of signal interference between devices operating close to each other. For this reason, any embedded device emitting RF signals must be compliant to specific international regulations before being available to the commercial market \cite{fcc}. 
On the other hand, unintentional magnetic emissions effectively represent a side-channel for any embedded device, providing useful information about the nature of the device and the specific activities executed at a given time instant \cite{Dejean2007}.

To provide few examples in this direction, recently the authors in \cite{camurati2018} provided a novel side-channel attack on mixed-signal chips, where digital logic circuits and the radio transceiver are coupled on the same chip. Using information leaked from digital circuits on a Nordic Semiconductor nRF52832 chip, the authors recovered the full AES 128-bit key, up to a distance of 10 meters from the target device. The authors in \cite{Cobb2012_tifs} and \cite{cobb2010} identified that the fabrication process of embedded circuits induced variations in the electrical properties of each chip. By using RF emissions from 16 different micro-controllers, they were able to identify uniquely each of them, recurring to imperfections of the fabrication processes. Similarly, the author in~\cite{wright2014} focused on the identification of the SCADA sensors and actuators in critical infrastructures and measured the unintentional emissions derived from the execution of a specific piece of code. Another application in this context is discussed in \cite{bihl2016}, where researchers used the unintentional RF emissions to achieve the classification and verification of IEEE 802.15.4 ZigBee wireless devices that are widely used in critical infrastructure applications. In the context of the IEEE 802.15.4 communication technology, unintentional RF emissions have been also investigated in \cite{Dubendorfer2012} and \cite{ramsey2012}, to build a system able to reject rogue devices. Concerning generic RF devices, the authors in~\cite{Suski2008} demonstrated the feasibility of fingerprinting wireless devices by looking at the non-idealities of the emitted RF signals. They showed that each packet emitted by a device has unique distinguishing features, that can be used to identify the transmitting entity. Indeed, we note that this fingerprinting methodology requires the transmission of RF packets, and thus, it cannot be applied to non-RF devices, such as the USB drives. Another use-case was introduced by the authors in \cite{lukacs2015}, where unintentional RF emissions and random noise waveforms were used to actively interrogate target microwave devices, recognize and classify antennas and terminations. Recently, the authors in~\cite{Cheng2019_ccs} used the magnetic signals radiated by the CPU to fingerprint heterogeneous devices, including laptops, smartphones, and additional mobile devices. This result has been achieved by precisely controlling the hardware, firmware, and software of the devices under tests, forcing the CPU to execute a stimulation software that enables devices fingerprinting.

A few works in the recent literature investigated the same context tackled by this contribution, i.e., malicious USB Flash Drives, with specific reference to the \emph{BadUSB} attack~\cite{Tian2016},~\cite{angel2016_usenix},~\cite{Tian2015}. For instance, the authors in~\cite{Tian2016} presented \emph{USBFILTER}, a software solution providing first packet-level access control for USB, and preventing unauthorized interfaces from successfully connecting to the host \ac{OS}. The authors in \cite{Griscioli2016} proposed an innovative approach, forcing the user to interact with the connected USB device before allowing the device to be used, thus ensuring that a real human-interface device is attached. The authors in \cite{angel2016_usenix} developed \emph{Cinch}, a countermeasure against malicious peripherals leveraging virtualization techniques to place the connected hardware in a logically separate machine, as an isolation layer from the main and protected one. This layer depends on security policies configured by the users to reject or accept interaction with connected USB peripherals. The authors in \cite{Tian2015} designed and implemented \emph{GoodUSB}, a mediation architecture for the Linux USB Stack. This solution defends against BadUSB attacks by enforcing permissions based on user expectations of device functionality. 
Recently, the authors in~\cite{suzaki2019} proposed to authenticate individual USB devices using tamper-proof Physical Unclonable Functions (PUFs), combining multiple security technologies available in commodity PCs, e.g., Trusted Platform Module (TPM), customized secure boot, and virtualization support. This is a software solution, that could not help in case the specific USB brand is a regular USB stick, such as in the Scenario described in our paper. We also report the interesting work by the authors in~\cite{bates2014}, where the firmware and software features of the USB protocol stack are used to identify the specific host.

On the one hand, we highlight that such solutions are specifically developed to thwart the BadUSB attack and its behavior \cite{badusbguide}. Today, the vendors of the devices used to carry out the BadUSB attack have fixed the vulnerabilities, thus making the BadUSB attack very hard to be realized. On the other hand, attackers are still able to cheat both the above defense measures and the current anti-malware software, e.g., by masquerading the malicious USB device as a legitimate one at the time of connection to the host \cite{Tian2016}. Especially in industrial companies and critical infrastructure scenarios, the impact of such attacks can be devastating, preventing the daily activities of the company and possibly leading to huge economic losses on the whole country relying on the target critical infrastructure.


Table~\ref{tab:comparison} summarizes the above contributions, along some reference system requirements.
\begin{table*}[htbp]
\caption{Qualitative comparison of \sol\ against competing solutions.}
\centering
\color{black}
    \begin{tabular}{|c|P{3.0cm}|P{2.2cm}|P{3.1cm}|P{2.cm}|}
    \hline
        \textbf{Ref.} & \textbf{No Host Firmware Modification} & \textbf{No Target Modification} & \textbf{Compatible with Read-Only Devices} & \textbf{No Radio Operations} \\ \hline
        \cite{camurati2018}     & \cmark & \xmark & \xmark & \cmark \\
        \cite{Cobb2012_tifs}    & \cmark & \xmark & \xmark & \cmark\\
        \cite{cobb2010}         & \cmark & \xmark & \xmark & \cmark\\
        \cite{wright2014}       & \cmark & \xmark & \xmark & \cmark\\
        \cite{bihl2016}         & \cmark & \xmark & \xmark & \cmark\\
        \cite{Dubendorfer2012}  & \cmark & \xmark & \xmark & \cmark\\
        \cite{ramsey2012}       & \cmark & \xmark & \xmark & \cmark\\
        \cite{Suski2008}        & \cmark & \cmark & \cmark & \xmark \\
        \cite{lukacs2015}       & \cmark & \xmark & \xmark & \cmark \\
        \cite{Cheng2019_ccs}    & \cmark & \xmark & \xmark & \cmark \\
        \cite{Tian2016}         & \xmark & \cmark & \cmark & \cmark \\
        \cite{angel2016_usenix} & \xmark & \cmark & \cmark & \cmark \\
        \cite{Tian2015}         & \xmark & \cmark & \cmark & \cmark \\
        \cite{Griscioli2016}    & \xmark & \xmark & \xmark & \cmark \\
        \cite{suzaki2019}       & \xmark & \xmark & \xmark & \cmark \\
        \cite{bates2014}        & \xmark & \xmark & \xmark & \cmark \\
        \hline
        \textbf{\sol}           & \cmark & \cmark & \cmark & \cmark\\
        \hline
    \end{tabular}
    \label{tab:comparison}
\end{table*}

Table~\ref{tab:comparison} highlights that a novel element characterizing \sol\ is its application to devices whose firmware and software is read-only, and therefore, not accessible by the verifier. Indeed, all the previous contributions on Radio Frequency (RF) fingerprinting require either the transmission of an RF packet by the device, e.g., an IEEE 802.11a packet in~\cite{Suski2008} and~\cite{Cobb2012_tifs}, or an identical sequence of operations on known data packets in~\cite{cobb2010}, or the access to the device under test, that have to execute a specific piece of code, e.g., the Ladder Logic Program (LLP) scan in~\cite{wright2014}. The first class of the above-described solutions restricts the application of fingerprinting to only devices featuring a radio. Conversely, the second class implies that the verifier has direct access to the firmware or the software of the device under test, by deploying a firmware/software update to allow for the fingerprinting process.
%
Conversely, our solution exploits the unintentional magnetic emissions of the boot process of a USB flash drive, therefore, enabling the fingerprinting process without requiring access to either the firmware or the software in the device under test---even when the device does not emit any RF signals. 
Further, our solution guarantees the full privacy of the device under test, making \sol\ overall enjoying the properties of being non-invasive, minimal-interactive, and privacy-preserving.

\section{Reference Scenarios and Adversary Model}
\label{sec:scenario_adv_model}

In the following, we assume two reference scenarios: (i) a private company; and, (ii) a critical infrastructure.
\\
{\bf Scenario \#1: Private Company.} We consider a private company, interested in maintaining very high levels of security and privacy for its own most important equipment and data. The employees of the company need to use USB flash drives for daily work, e.g., to transfer files between different equipment. 
Being aware of the vulnerabilities associated with the USB standard specification, the system administrator would like to deploy a solution enabling the connection of authorized USB devices only, i.e., brands and models that are not affected by any known vulnerability.
Therefore, the company provides to the employees its own USB flash drives, characterized by a specific brand and model. Only the USB flash drives of this particular brand and model can be used by the employees, while the use of any other brand or model of USB flash drive is not allowed.
To enforce such a protection strategy, the IT system administrator deploys a host device on purpose, dedicated to the measurement and testing of specific USB Flash Drives brought into the company by the employees. By connecting the USB Flash Drives to the host, the system administrator could verify that the brand and the model of the USB Flash Drive are effectively the ones authorized by the company. It is worth noting that the physical deployment of such a host device can be optimized to be located within an \emph{electromagnetic safe zone}, where the electromagnetic interference due to other devices is limited.
\\
{\bf Scenario \#2: Critical Infrastructure.} We also consider a digitized Critical Infrastructure, such as the control center of an airport, a smart port, an electricity control station, a smart grid, or a train station, to name a few. Being very sensitive computing centers, the computational units behind these critical infrastructures are usually highly protected, and equipped with minimal interfacing capabilities toward the public Internet. As such, the usage of USB Flash Drives is allowed only for unavoidable tasks, such as the installation of critical updates to the system, or the export of data stored locally.
Indeed, such critical infrastructures can be the target of malicious attackers, aiming at compromising the operation of a whole country by shutting down, e.g., its transportation or electricity network. Thus, the computing equipment controlling the operation of these critical infrastructures need to be carefully protected against any hazard, including the usage of malicious USB Flash Drives. 
To enforce such a protection strategy, we assume that the system administrator would like to use only a specific USB Flash Drive, and reject any other one, even if it is of the same brand and model. To this aim, the system administrator could deploy a dedicated host device for the test of the USB flash drive, and to deploy a system enabling and supporting the unique identification of the specific USB Flash Drive. As for the previous scenario, such a host device can be deployed where electromagnetic interference is minimal, to optimize the overall measurement procedure.
\\
{\bf Adversary Model.}  We assume an adversary that is interested in compromising the equipment of the company or the Critical Infrastructure, by adopting malicious USB flash drives. The specific aims of the adversary could be manifold, e.g., injecting malware into the ICT infrastructure where the USB flash drives are connected to, redirecting the Internet traffic to malicious websites, or stealing information stored locally in the ICT infrastructure, to name a few.

To this aim, the adversary can replace the regular (allowed) USB flash drive with another USB flash drive, having the same external look and form factor of the previous one, but a different hardware and/or firmware. This USB flash drive is modified on purpose to achieve one of the objectives listed above. 
In addition, we also assume that the adversary can exploit weaknesses in the USB firmware deployment and replace the legacy firmware of the USB flash drive provided by the company, injecting its malicious version \cite{badusbguide}.  
It is worth noting that in such a scenario, the deployment of regular anti-malware software would not be effective. Indeed, as highlighted by previous contributions \cite{Nissim2017}, attacks exploiting USB external drives can modify the firmware of the component in such a way to appear to host devices as regular USB equipment, e.g. a keyboard. Thus, the anti-malware would recognize the keyboard as a legitimate component and it would grant the related permissions.

We also assume that the adversary does not have access to the source code of the firmware deployed in a specific USB flash drive. This is confirmed by the fact that the firmware on-board of commercial USB flash drives cannot be changed and even cannot be accessed by the end-users after the deployment. All the hacks that are described in technical and unofficial blogs refer to particular firmware versions of specific USB flash drives\footnote{\url{https://arstechnica.com/information-technology/2014/07/this-thumbdrive-hacks-computers-badusb-exploit-makes-devices-turn-evil/}} \footnote{\url{https://www.reddit.com/r/netsec/comments/112kuv/reprogram_usb_flash_drive_microprocessors_firmware/}}. However, the manufacturers immediately withdrew the affected devices from the market, and they updated the native firmware to fix these weaknesses. Therefore, we believe that in the vast majority of practical cases the native firmware of commercial USB flash drives cannot be modified by end-users.

Even in the unlikely case where an adversary can access the firmware, we consider it extremely hard and impractical to reverse-engineer it to obtain the source code, and to re-deploy it on the same USB flash drive. Generally, the firmware is protected by intellectual property rights, the source code of such firmware is secret, and protected by multiple security layers deployed at manufacturing time, and not available for public download.

Finally, we notice that the two scenarios described above always assume an end-user that could not circumvent the deployed system, e.g., by not testing its USB flash drive before usage. This assumption is consistent with the scenario of a benign employee, not aware of the (possible) replacement of its USB Flash drive with a malicious one.

\section{The \sol\ Framework}
\label{sec:idea}

\begin{figure}[htbp!]
    \includegraphics[width=.5\columnwidth]{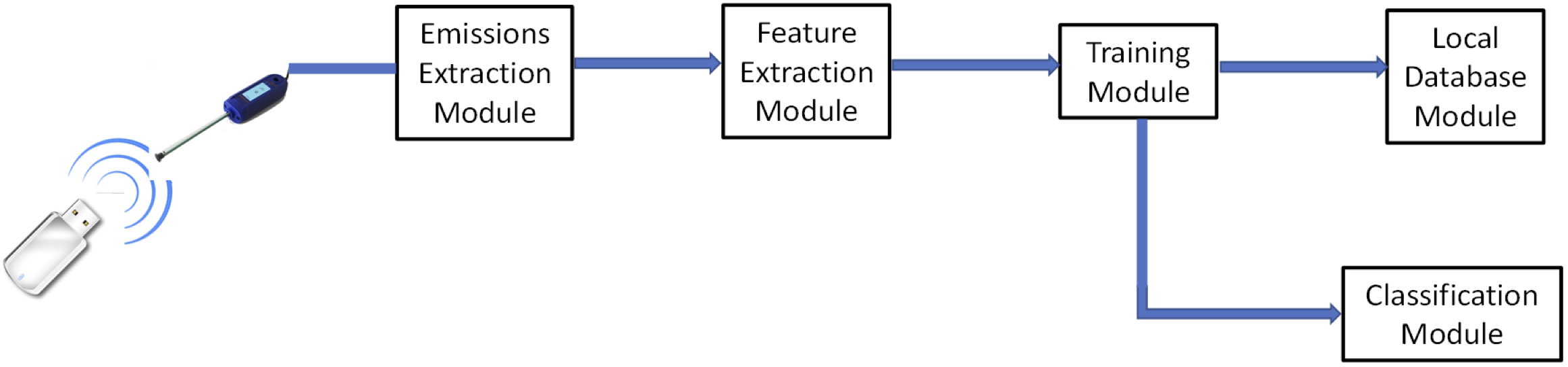}
    \centering
    \caption{Logical architecture of the \sol\ framework.}
    \label{fig:sys}
\end{figure}

\sol\ consists of five different modules, as depicted in Figure \ref{fig:sys}.

We identify two modes of operation:
\begin{itemize}
    \item \emph{Training Mode.} During this phase, a local database is populated with all the profiles of the available USB devices, including the authorized ones.
    \item \emph{Classification Mode.} This is the online operational mode of \sol, where a new unknown USB device is tested to verify if its profile matches the one already collected during the training mode.
\end{itemize}

The details of the modules involved in the \sol\ framework are provided below:

\begin{itemize}
    \item[$\bullet$] \emph{Emissions Extraction Module.} The role of this module is to detect, capture, and log unintentional magnetic emissions from a particular device. Specifically, \sol\ focuses on the analysis of the unintentional magnetic emissions leaked by USB flash drives at boot time, i.e., when the USB device is first connected to the host device and it executes boot operations.  Then, the raw data consisting of: (i) a timestamp; (ii) the acquisition frequency; and, finally, (iii) the value of the \ac{RSS} are passed to the Features Extraction Module. 
    The operations of the Emissions Extraction Module are carried out with particular hardware equipment, able to capture magnetic emissions on the specific operating frequency thanks to a suitable magnetic antenna. More details about the equipment used in our experiments will be provided in Section \ref{sec:experiments}. 
    \item[$\bullet$] \emph{Features Extraction Module}. Starting from the raw data provided by the Emissions Extraction Module, this module is responsible for generating the features of interest for the particular signal. Specifically, this module operates in three different phases, i.e., \emph{Data Normalization}, \emph{Regions Definitions}, and \emph{Features Computation}.
    \begin{itemize}
        \item[-] \emph{Data Normalization}. The data acquired by the Emissions Extraction Module are normalized by re-centering them to the minimum value of the dynamic range of the observation, and then, by re-scaling them by the value of their dynamic range. Specifically, assuming $x_i$ is a sample of the distribution, and $X_{MIN}$ and $X_{MAX}$ are the minimum and the maximum value of the dynamic range, the normalized sample $\hat{x_i}$ is computed as: $\hat{x_i} = \frac{x_i - X_{MIN}}{ \left( X_{MAX} - X_{MIN} \right) } $. 
        
        These operations are crucial to enable cross-comparisons between different measurements, as they contribute to eliminate any small difference in the absolute value of the received power---e.g., due to small misalignment of the measurement setup.
        \item[-] \emph{Regions Definition}. Each value of the \ac{RSS} reported by the \emph{Emissions Extraction Module} is associated with a given timestamp and a specific frequency. In this phase, the \emph{Features Extraction Module} merges in the same observation vector the measurements within a given time and frequency regions, that can be configured uniquely for any specific investigation. The output of this phase is a matrix, containing for each identified region all the collected \ac{RSS} values for the particular experiment.
        \item[-] \emph{Features Computation}. Starting from the matrix created in the previous phase, the \emph{Features Computation} phase is responsible for computing the features on each observation vector.
        
        Assuming now that $X = [x_1, x_2, ..., x_i, ..., x_N]$ is the observation vector containing the normalized intensity of the acquired signal in a given time frame and frequency range, we consider the following statistics: (i) mean; (ii) standard deviation; (iii) variance; (iv) skewness; and, finally, (v) kurtosis, to precisely characterize the unintentional radiated emissions from each USB device. Equations \ref{eq:skewness} and \ref{eq:kurtosis} report the formulas for the skewness and kurtosis, respectively, where $(\bar{X})$ is assumed to be the mean value of the observation vector.
        \begin{equation}
            \label{eq:skewness}
            \gamma =\frac{\tfrac{1}{N} \sum_{i=1}^N (x_i-\bar{X})^3}{\left[\tfrac{1}{N-1} \sum_{i=1}^N (x_i-\bar{X})^2\right]^{3/2}}.
        \end{equation}
        
        \begin{equation}
            \label{eq:kurtosis}
            \kappa = \frac{\sum_{n=1}^N \left( x_i - \bar{X} \right)^4} {\frac{1}{N} \left( \sum_{n=1}^N \left( x_i - \bar{X} \right)^2 \right)^2  }.
        \end{equation}
        
        The output of this phase is a new matrix, containing the value of each feature for the reference experiment.
    \end{itemize}

    The current matrix containing the values of the features for each time and frequency interval can be passed either to the Training Module or the Classification Module, according to the particular operating mode of \sol.
    \item[$\bullet$] \emph{Training Module.} This module, involved only in the Training Mode, is responsible for creating the reference profile of the particular device under investigation, by using the features previously identified by the \emph{Features Extraction Module}. The created profile is then uploaded to the \emph{Local Database Module}.
    \item[$\bullet$] \emph{Local Database Module.} The local database stores information about the specific training model associated with each device. When a new (previously unknown) device is registered, the \emph{Training Module} establishes the specific statistical model for this device, that is stored in the \emph{Local Database Module}. At the same time, when the \emph{Classification Module} is executed, the \emph{Local Database module} provides the reference statistical models used for comparison and classification. 
    \item[$\bullet$] \emph{Classification Module.} This module, involved only in the Classification Mode, is in charge of establishing if the previously stored profiles match the one extracted from the device under test. The comparison is carried out by considering the specific features considered by the \emph{Features Extraction Module}, and by comparing the obtained features with the stored profiles available in the \emph{Local Database Module}. The module outputs a similarity score, representing how much the magnetic emissions collected from the device under test are similar to the stored profile. If such a score is equal or greater than zero, the device is assumed to be authentic. Otherwise, it is discarded. 
    Among the several possible classifiers available in the literature \cite{Shabtai2009}, in this paper, we have adopted the one-class \ac{SVM} classifier, since it natively fits our objectives. More details will be provided in Section \ref{sec:experiments}.  
    
\end{itemize}

Using the five modules described above, depending on the scenario and the specific objective, either the brand and the model of the USB Flash Drive or the specific USB device can be uniquely characterized when connected to a host device. 

Wrapping up, during the \emph{Training Mode}, the unintentional magnetic emissions leaked by the USB devices are collected via the hardware equipment and the functionalities offered by the \emph{Emissions Extraction Module}. Then, the reference features are extracted from the raw signals by the \emph{Features Extraction Module}, and a reference statistical model is created by the \emph{Training Module} and stored into the \emph{Local Database Module} for future use. 
The process is performed for a large number of devices, to create a database containing an exhaustive number of products of the same device.

During the \emph{classification mode}, all the USB devices are tested, by connecting them to a reference host device and collecting the unintentional magnetic emissions during the boot procedures via the functionalities offered by the \emph{Emissions Extraction Module}. The reference features are computed from the acquired signal, and then passed to the Classification Module, which provides a similarity score, indicating how much the stored profile matches the provided one. 
If no correspondence is found (i.e., the similarity score is negative), the new device is assumed as unauthorized. Instead, if the profile of the unintentional magnetic emissions is found to be consistent with the allowed class (i.e., the score is equal to 0 or positive), the device will be authorized.

It is worth noting that \sol\ focuses on the boot procedures of USB devices, because they are strictly required for the normal operations of the USB themselves. Indeed, being loaded in the firmware, the boot procedures associated with the USB device should not be changed by a non-malicious user. Otherwise,  any modification to the firmware (either intentional or not) will generate significant changes to the magnetic emissions during the boot procedures, indicating therefore that the USB flash drive has been either tampered with or replaced. Such modifications of the behavior of the USB flash drive during the boot operations can be detected by \sol\ and can lead to a timely rejection of the device. More details on the specific modifications that can be done when tampering with USB devices are provided in Section~\ref{sec:discussion}.

\section{Experimental Assessment}
\label{sec:experiments}

In this section, we provide the details of our experimental campaign, aimed at measuring the performance of \sol\ for the classification of either the brand and model of USB flash drives, or the specific USB flash drive, according to the scenarios previously introduced in Section \ref{sec:scenario_adv_model}.
Section \ref{subsec:tools} illustrates the equipment used in our experiments, Section \ref{sec:consistency} provides evidence of the consistency among different observations of boot operations on the same USB flash drives, Section \ref{sec:class_usb} shows the results achieved by \sol\ in the classification of different brand and models of USB flash drives, Section \ref{sec:auth} illustrates how \sol\ achieves the identification of single USB flash drives using more advanced equipment, Section~\ref{sec:exp_fr} provides the accuracy of \sol\ when a reduced number of features is selected, Section~\ref{sec:fw_mod} includes some experiments on firmware modification and, finally, Section~\ref{sec:host_impact} illustrates how the features of the host device impact on the profile of unintentional magnetic emissions radiated by USB flash drives. \\
The code of \sol, including the source data of our experiments and the tools necessary to reproduce our results, are available for download at \cite{crilab}.

\subsection{Tools and Methodology}
\label{subsec:tools}

In our experimental campaign, we used the equipment listed below.
\begin{itemize}
\item \textbf{USB Devices}. We tested the performance of \sol\ with a set of total of 59 different USB devices, related to 17 unique models, whose details are reported in Tab. \ref{tab:usblist}, including the model of the included \emph{Controller Chip}, containing the local oscillator (mainly) responsible for magnetic emissions generation (we notice that the manufacturer SanDisk does not provide any detail about the controller chip).
\begin{table*}[htbp]
\centering
\color{black}
\caption{List of USB Brands and Models adopted in our experiments.}
\begin{tabular}{|l|l|l|}
\hline
\textbf{Device} & \textbf{Controller Chip} & \textbf{Device ID} \\ \hline
HPx900w-64GB \cite{HP}& PHISON-PS2251-09-V & U1 \\ \hline 
JUANWE-32GB \cite{JU} & FirstChip-FC1179 & U2 \\ \hline
 Kingston Digital 16GB Data Traveler G4 \cite{KD} & PHISON-PS2251-09-V & U3 \\ \hline
Mosdart-8GB \cite{MD} & AlcorMP & U4 \\ \hline
PNY Turbo 128GB \cite{PNY} & PHISON-PS2251-09 & U5 \\ \hline
Patriot 128GB Supersonic Rage Series \cite{PR} & PHISON-PS2251-09-V & U6 \\ \hline
Rubber Ducky \cite{RubberDucky} & Atmel 32UC3B1 & U7 \\ \hline
Samsung BAR Plus 32GB \cite{SB} & Silicon Motion SM3267 & U8 \\ \hline
SanDisk Ultra 128GB \cite{SU}& SanDisk & U9 \\ \hline
SanDisk Cruzer 128GB \cite{S128} & SanDisk & U10 \\ \hline
SanDisk Cruzer 32GB \cite{S32} & SanDisk & U11 \\ \hline
SanDisk Cruzer 64GB \cite{SD} & SanDisk & U12 \\ \hline
Silicon Power 32 GB Blaze B30 \cite{SP} & PHISON-PS2251-09-26 & U13 \\ \hline
Toshiba TransMemory 64GB \cite{TO} & PHISON-PS2251-11 & U14 \\ \hline
SanDisk 16 GB \cite{S16GB} & SanDisk  & U15 \\ \hline
Strontium 16 GB \cite{StrontiumUSB} & Strontium  & U16 \\ \hline
Klevv Neo C20 16GB \cite{KlevvNeo} & Essencore & U17 \\ \hline

\end{tabular}
\label{tab:usblist}
\end{table*}
We selected these brands to create a large dataset, considering both the most used models and any possible factor affecting the magnetic emissions of such devices. Specifically, SanDisk, Samsung, and Kingston USB flash drives are the top three brands purchased by users on Amazon---as of July 2019 \cite{AmazonList}. At the same time, Juanwe, PNY, HP, and Mosdart are all positioned among the top 10 brands. 
In addition, we also selected some brands based on the use of the same version of the micro-controller. For instance, we highlight that the micro-controller of the HP (U1), the Kingston (U3), the PNY (U5), the Patriot (U6), the Silicon Power (U13), and the Toshiba (U14) are the same, i.e., they feature the same hardware chip, while the layout of the \ac{PCB} and/or the versions of the firmware are different (e.g., 09-V for the Kingston, 09-26 for the Silicon Power) \cite{phison}. Also, it is also worth noticing that we considered different devices having the same brand, model, and firmware: specifically, we considered $15$ \emph{SanDisk 16GB}, $15$ \emph{Strontium 16GB} and $15$ \emph{Klevv Neo C20 16GB}.
All the devices have been analyzed as provided by the manufacturer, without any modification to the legacy firmware. To prove the effectiveness of our method, we acquired the magnetic emissions by temporarily removing the case of the USB flash drive. 
Besides, we highlight the presence of a \emph{malicious} device, i.e., a USB Rubber Ducky. This USB Flash drive is a cross-platform keystroke injection tool, able to work effectively with either Windows, Mac, or Linux. It features a 60 MHz 32-bit processor, hidden in the form factor of an ordinary generic USB flash drive. It consists mainly of a CPU chip and a micro SD storage, used to upload malicious payloads to the host device. 
Once connected to a host machine, the Rubber Ducky is recognized as a keyboard. Thus, the host device accepts its payload, consisting of pre-programmed regular keyboard strokes, up to a maximum rate of 1000 words per minute \cite{RubberDucky}. The commands are programmed using a dedicated scripting language, and it is possible to perform a wide range of automated functions, such as password brute-forcing, binaries injection, shell reversing, as well as changing system settings. Nowadays, it is mainly used by professionals, penetration testers, and system administrators.
    
\item \textbf{Aaronia PBS2 EMC Probe set}. To collect the unintentional magnetic emissions radiated by the USB devices, we used the Aaronia PBS2 EMC Probe set \cite{PBS2}. This equipment allows straightforward pinpointing and measurement of interference in the frequency band from the DC (1Hz) up to 9GHz in electronic component groups, as well as the execution and monitoring of generic Electro-Magnetic measurement. 
As a probe, we used a 25mm magnetic (H) field probe PBS-H3, covered with an insulating layer, thus providing a safe measurement environment for oscillators and mains lines. This probe is connected to the UBBV2 40dB EMC RF pre-amplifier, allowing for a clear separation between the useful signal and the surrounding noise. The probe is then connected via a low-impedance cable to a \ac{SDR}.
    
\item \textbf{HackRFOne \acl{SDR}}. The HackRFOne \ac{SDR} has been used to record unintentional magnetic emissions acquired from the Probe Set over a frequency span of 10 MHz, and to convert raw data (in the form of regular I-Q samples) into spectral power density measurements. Specifically, the SDR performs a \ac{FFT} on the provided I-Q data and it generates, for each time frame, a tuple containing a reference timestamp (in milliseconds), a frequency (in Hz), and a power level (in dBm). Tab. \ref{tab:hackrf_conf} provides the details of the configuration of the \ac{SDR}.
\begin{table}[htbp]
\centering
\caption{Configuration of the HackRFOne Software Defined Radio.}
\begin{tabular}{|P{4.8cm}|P{2.8cm}|}
    \hline
    \textbf{Feature} & \textbf{Value} \\ \hline
        \text{Resolution Bandwidth} & 976.6 Hz \\ \hline
    \text{Center Frequency} & 5 MHz \\ \hline
    \text{Start Frequency} & 0 MHz \\ \hline
    \text{Stop Frequency} & 10 MHz \\ \hline
    \text{Reference Level} & -5 dB \\ \hline
    \text{FFT Size} & 8192 samples \\ \hline
\end{tabular}
\label{tab:hackrf_conf}
\end{table}
    
\item \textbf{Rohde \& Schwarz FSW8 Spectrum Analyzer}. The Rohde \& Schwarz FSW8 Spectrum Analyzer has been used in place of the HackRFOne to record unintentional magnetic emissions acquired from the Probe Set over a larger frequency span, up to 200 MHz. As with the HackRFOne, the equipment automatically converts raw data into spectral power density measurements, by performing the \ac{FFT} over the acquired samples, and it provides a tuple for each time frame, containing a timestamp (in milliseconds), a frequency (in Hz), and a power level (in dBm). Table~\ref{tab:SAconf} provides the details of the configuration of the spectrum analyzer.
    
\begin{table}[htbp]
    \centering
    \caption{Configuration of the Rohde \& Schwarz FSW8 Spectrum Analyzer.}
    \label{tab:SAconf}
    \begin{tabular}{|P{4.8cm}|P{2.8cm}|}
    \hline
    \textbf{Feature} & \textbf{Value} \\ \hline
    Video Bandwidth & 3 MHz \\ \hline
    Resolution Bandwidth & 3 MHz \\ \hline
    Center Frequency & 100 MHz \\ \hline
    Start Frequency & 0 MHz \\ \hline
    Stop Frequency & 200 MHz \\ \hline
    Frequency Span & 200 MHz \\ \hline
    RF Attenuation & 10 dB \\ \hline
    Reference Level & -5 dB \\ \hline
    Sweep time & 4.01 ms \\ \hline
    Sweep points & 4001 \\ \hline
    \end{tabular}
\end{table}

\item \textbf{Matlab R2020a}. Matlab R2020a has been used to implement the \emph{Features Extraction Module}, the \emph{Training Module}, the \emph{Local Database Module}, and the \emph{Classifier Module}. As the classification algorithm, we used the one-class \ac{SVM} classifier provided by the Machine Learning Toolbox of Matlab. Specifically, the one-class SVM classifier analyzes each of the USBs as a standalone USB flash drive, by creating a profile that matches the features provided for the particular USB brand (or the specific USB flash drive). When a test set is provided in input for classification, as an output, the one-class SVM classifier provides an \emph{evaluation score} for each input sample. Such evaluation score indicates the likelihood that the specific input sample is consistent with the created model, or not. This approach is particularly useful when the range of possible classes is wide, such as for USB flash drives, and thus a multi-class classification approach is not suitable.
\end{itemize}

\begin{figure}[htbp]
    \includegraphics[width=.5\columnwidth]{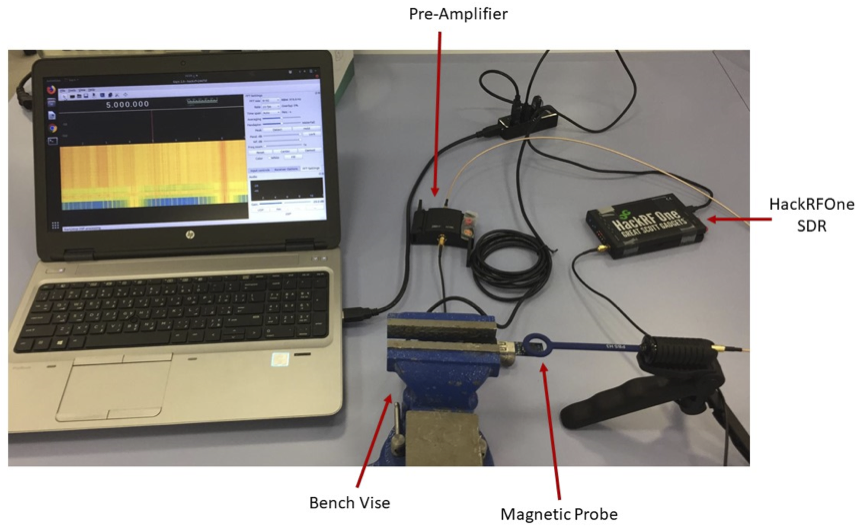}
    \centering
    \caption{Measurements Setup.}
    \label{fig:setup}
\end{figure}

All the experiments described in the following subsection have been conducted by collecting the unintentional magnetic emissions radiated by the target USB devices in regular laboratory conditions, without any effort to reduce the environmental noise. Our measurement setup is shown in Figure \ref{fig:setup}. 

Note that a USB cable extension was placed on a Bench Vise, to physically stabilize the USB device under test. In addition, this setup allows having uniform conditions for emissions collection. The Magnetic Probe has been placed on top of the USB device controller chip, in a way to clearly capture the radiated emissions. 
For each USB device under test, ten different boot sequences were acquired, where each measurement was at least 3.35 seconds long.

Finally, for the data processing, we used a Lenovo Ideapad 320 laptop, equipped with an Intel i7-7500 processor running at 2.70 GHz and equipped with 8 GB of RAM.

\subsection{Consistency of Unintentional Magnetic Emissions Radiated from USB Devices}
\label{sec:consistency}

Our first investigation aims at establishing the feasibility of uniquely identifying either the brand and model of the USB flash drive or the specific USB device, based on the profile of the unintentional magnetic emissions.

\begin{figure}[htbp]
    \includegraphics[width=.49\columnwidth]{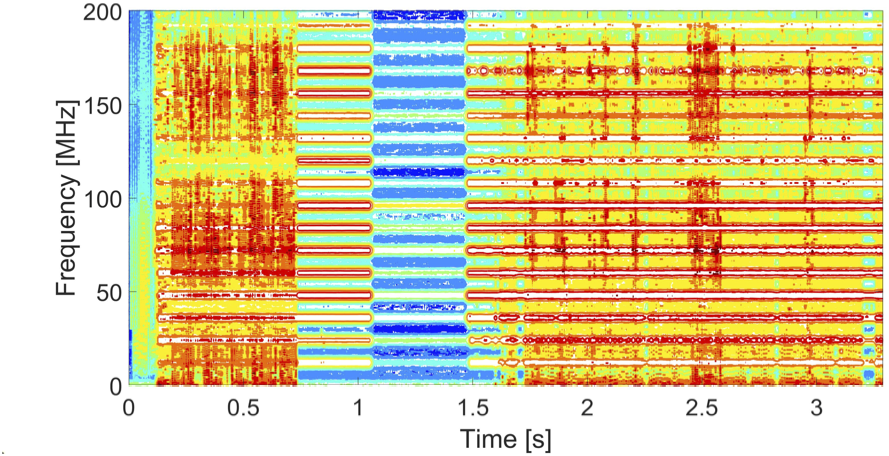}
    \includegraphics[width=.49\columnwidth]{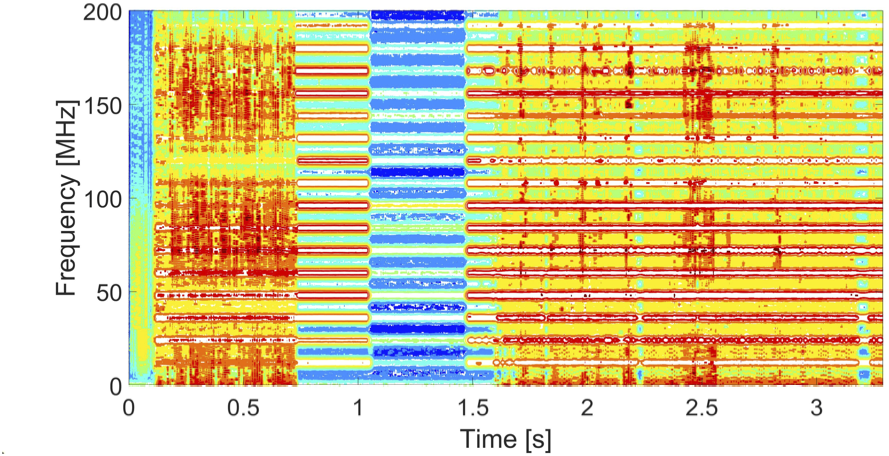}
    \centering
    \caption{Profile of the unintentional magnetic emissions of a Rubber Ducky USB flash drive (U7), during two different executions of the boot procedures.}
    \label{fig:RD}
\end{figure}

Figure \ref{fig:RD} shows two sample measurements acquired at different times, related to the unintentional magnetic emissions radiated from the same Rubber Ducky USB flash drive (U7), during the boot procedures. Because of the normalization phase occurring in the \emph{Features Extraction module}, all the samples corresponding to given times and frequencies have a value between 0 and 1. Specifically, the blue color maps values in the range $\left[ 0 - 0.25 \right[$, the cyan corresponds to values in the range $\left[ 0.25 - 0.5 \right[$, the yellow is related to values in the range $\left[ 0.5 - 0.75 \right[$, while the red color indicates value in the range $\left[ 0.75 - 1 \right]$.

First, we notice that the time duration of our experiments is enough to capture not only the boot phase, where the whole system is powered up and initialized, but also to acquire the following idle state, where the USB device waits for instructions by the target USB device.

By comparing the two figures, it emerges that different recordings have similar normalized power profiles, as per the time and the frequency. Indeed, small differences can be found in the time domain, due to a different delay between the time instant at which the acquisition was started and the time instant where the USB device was connected to the host device, and in the intensity of the unintentional radiation emissions, due to the surrounding noise and minimal displacement between the USB device and the probe. However, even across different measurements, different acquisitions of the unintentional radiated emissions from the same USB device are consistent, confirming that their statistical analysis can be effectively used for discriminating the USB brand and model. 


\subsection{Scenario \#1: Classification of USB flash drives brand and model}
\label{sec:class_usb}

The \emph{Scenario \#1} described in Section \ref{sec:scenario_adv_model} involves the identification of the brand and model of the specific USB Flash Drive under test. As shown in the previous section, different brands of USB Flash Drives are characterized by a very different profile of magnetic emissions. This motivates the use of relatively cheap equipment, i.e., the HackRFOne \ac{SDR}, as the tool for data acquisition and processing. We recall that the HackRFOne has a maximum acquisition bandwidth of 10 MHz. Thus, this frequency span can be used only to classify USB flash drives brands and models.

To enable cross-comparisons, we considered a fixed observation window of 1.08 seconds for each of the acquired traces. This time-slot was further divided into a given number of time and frequency regions, leading to a specific number of features characterizing each observation.

To provide sufficient robustness to the classification, we considered a total number of $325$ features, computed as follows. We first considered the overall observation interval of 1.08 seconds, and we computed (5) features, i.e., the mean, standard deviation, variance, skewness, and kurtosis of the samples in the whole interval (see Section~\ref{sec:idea} for the details of the computation of each feature). This step results in a total of 5 features. Then, we further divided the overall observation interval of 1.08 seconds in four (4) time regions, each region lasting $0.27$~s. For each time region, we computed the same five (5) features listed above. This step results in a total of 20 features. Then, we further divided each of the four (4) time regions into fifteen (15) frequency regions, where each frequency region has a bandwidth of $666.67$~Hz. For each of the $15 \cdot 4$ regions, we computed the same five (5) features listed above. This step results in a total $15 \cdot 4 \cdot 5 = 300$ features. Summing up the three steps, we have a total of $5 + 20 + 300 = 325$ features.

Specifically, the detection of authorized devices against not-authorized ones has been performed resorting to the one-class SVM classifier (see Section~\ref{sec:idea} for more details). For the Scenario \# 1, we considered 17 different models of USB Flash Drives (classes), i.e., \{$U1, \ldots, U17$\}, and we performed $10$ observations (measurements) for each class, for a total of $170$ observations. Moreover, we adopted the 10-fold cross-validation where, in each fold, $90\%$ of the observations (from each class) have been selected for the training process. Then, we test the remaining observations ($10\%$) against the models created for each USB flash drive (disjoint from the test set and therefore not containing any testing sample), and we evaluate the accuracy of the model through a \emph{similarity score}. Note that we have $17$ \emph{similarity scores} for each testing sample, i.e., the score resulting from the evaluation of the testing sample against each of the $17$ classes (USB flash drives). Therefore, this strategy leads to a set of $17$ similarity scores for each testing sample, and a total number of $17 \cdot 170 = 2,890$ \emph{similarity scores}.

\begin{figure}[htbp]
    \includegraphics[width=.6\columnwidth]{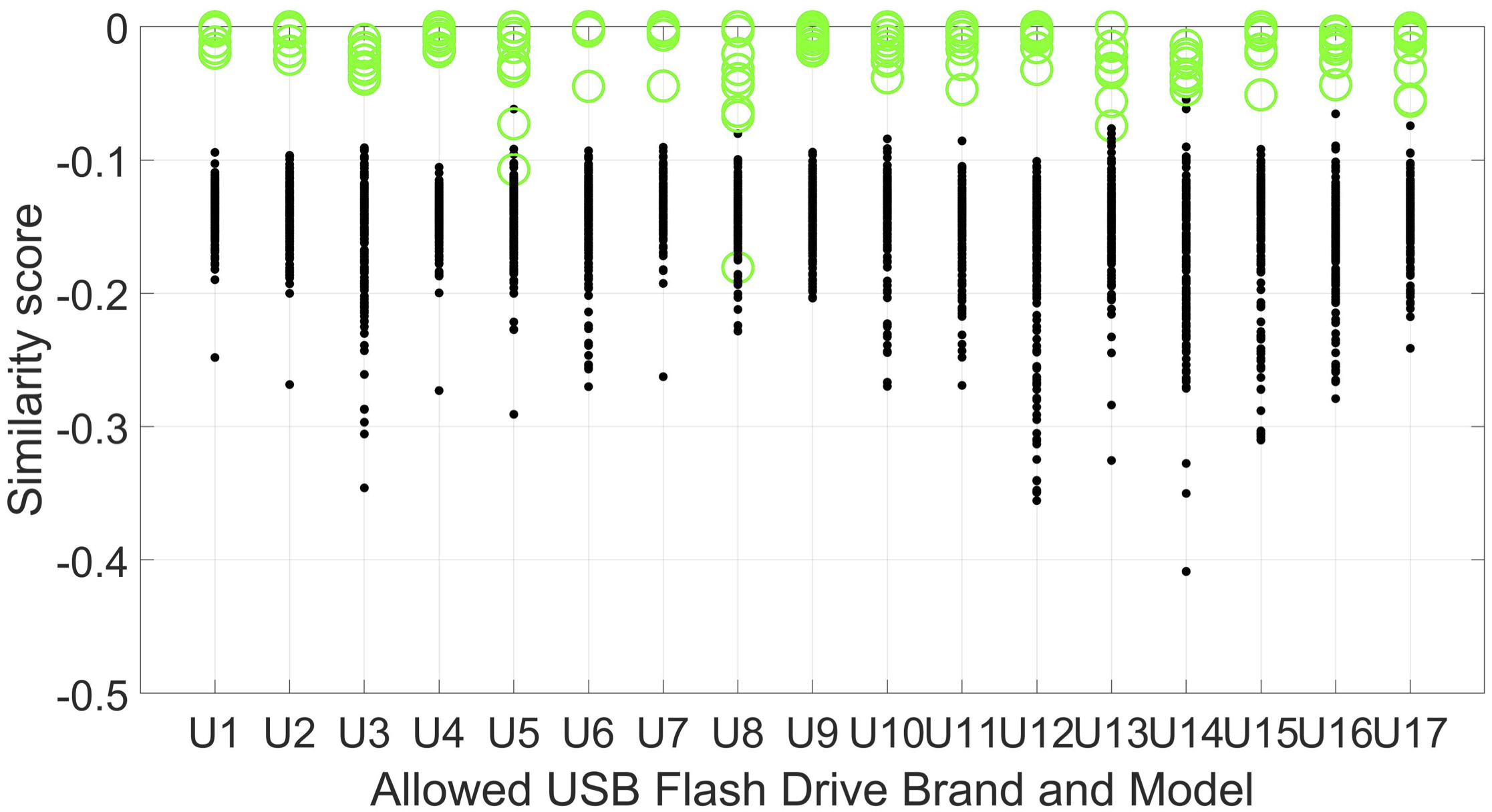}
    \centering
    \caption{Classification performance of \sol. Each column refers to a set of $170$ experiments where just one USB flash drive model is assumed to be authorized. Green circles report the value of the similarity score of the classifier for the authorized device, while black dots report similarity score for the unauthorized USB brands.}
    \label{fig:classification_scenario1}
\end{figure}

Figure~\ref{fig:classification_scenario1} shows the computed similarity scores as a function of the 17 considered classes \{$U1, \ldots, U17$\}. For each class (USB device), we reported as green circles ($10$ samples per class) the observations coming from the authorized device, while we considered the black dots ($160$ samples) for the non-authorized ones. 

It is worth noting that, in the vast majority of the experiments, the authorized USB flash drives report values that are very close to $0$ and less than the value of $-0.01$, while the unauthorized ones report lower scores. Overall, given that the model of each USB flash drive is trained and validated on its own, we should select a threshold value for each USB brand, and decide to \emph{accept} it as authorized if the evaluation score is higher than the threshold. As a worst-case, we notice that selecting a general threshold $T=-0.075$ guarantees \ac{TPR} $0.982$ (i.e., $98$\% of the authorized brands are correctly recognized), and a \ac{FPR} of only $0.001$ (i.e., only $0.01$\% of the unauthorized brands are erroneously recognized as authorized). However, a threshold can be selected for each USB flash drive. For instance, focusing on the USB Flash Drive $U7$, the threshold value $T_7 = -0.0365$ provides TPR $1$ and FPR $0$, i.e., the best possible result.\\
We remark that the above results show that \sol\ can distinguish the brand and the model of the USB flash drive, even if the brands share the same controller chip version and same PCB layout. For instance, we can distinguish the specific USB flash drive model among the HPx900w-64GB, the Kingston Digital 16GB Data Traveler G4, and the Patriot 128GB Supersonic Rage Series, despite these three models share the same controller chip and PCB layout, i.e., the PHISON-PS2251-09-V.\\
The above results suggested that, to some extent, \sol\ can identify not only the specific hardware chip but also the code that is executed during the boot. 
Indeed, \sol\ can discriminate also between the HP, Kingston, PNY, Patriot, Silicon Power, and Toshiba USB flash drives, even if they are based on the same hardware chip. Indeed, we recall that their PCB and/or the version of the firmware is different, and this leads to a different profile in the unintentional magnetic emissions at lower frequencies during the boot operations. More details on the impact of the firmware running on a particular chip will be provided in Section~\ref{sec:fw_mod}.

We also investigated the time required by \sol\ to provide the decision. Using 325 features, over a set of $2,890$ experiments, \sol\ requires $11.5$~ms on average, with a lower confidence interval of $10.27$~ms and higher confidence interval of $12.72$~ms.

Finally, we notice that, in principle, any Machine-Learning (ML) and Deep Learning (DL) classification tool can be used to precisely classify the magnetic emissions radiated by the tested USB Flash Drives. However, we highlight that the scope of our paper is not to find the best classification tool for this problem, but to show that unintentional magnetic emissions can be used to fingerprint the brand and the model of commodity USB Flash Drives. The data acquired from the tested USB Flash Drives have been released as open-source, to give to interested researchers the opportunity to test additional ML or DL classification tools and further boost the performance of \sol \cite{crilab}.

\subsection{Scenario 2: Identifying the specific USB Flash Drive}
\label{sec:auth}

In the context of the \emph{Scenario \#2}, it is of crucial importance to uniquely identify the specific USB flash drive, to protect the Critical Infrastructure against malicious software injection. As shown in Section \ref{sec:consistency}, identifying uniquely a USB Flash Drive involves discriminating very subtle differences in the profile of the unintentional magnetic emissions, possibly spanning a wide spectrum. To this aim, more powerful equipment should be used, characterized by a wider real-time bandwidth resolution, such as the \emph{Rohde \& Schwarz FSW8} Spectrum Analyzer, leading to an increased number of features, that could be processed by a more powerful processing unit.

To investigate the feasibility of this setup of \sol\ to uniquely identify USB Flash Drives, we considered USB Flash Drives of the same brand and model, and we applied \sol\ to identify them uniquely. 
We considered a fixed observation window of 3.35 seconds for each of the acquired traces, and the same number of features ($325$), derived as previously explained.
Note that the methodology to test the performance of \sol\ is still the same previously described: in each experiment, we assumed a selected USB Flash Drive to be the allowed one, while the others were assumed to be rejected, and we used the 10-fold cross-validation, where in each fold, 90\% of the recordings were used in the \emph{Training Mode}, and the remaining 10\% were used to test the \emph{Classification Mode}.

\begin{figure}[htbp]
    \includegraphics[width=.6\columnwidth]{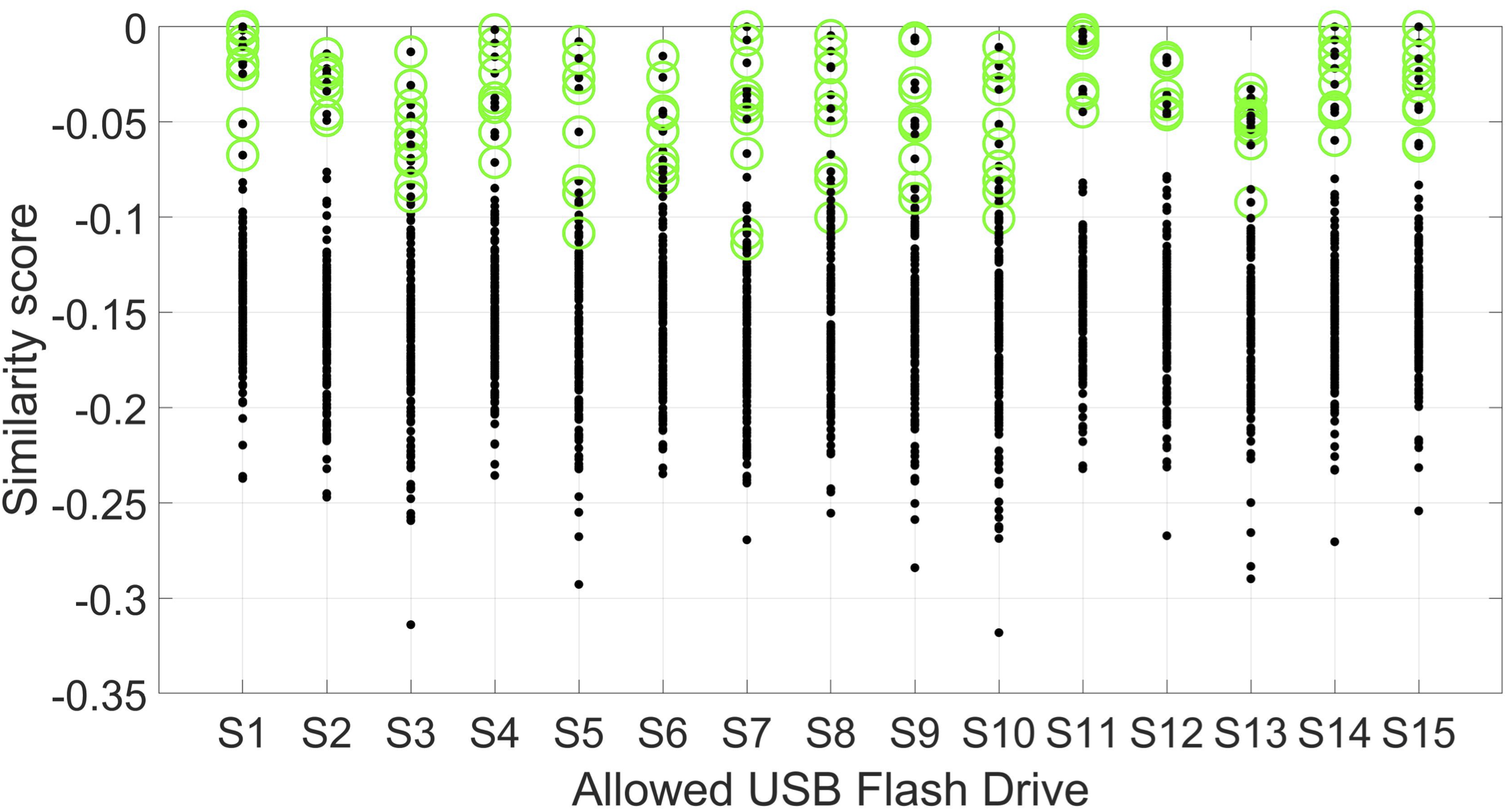}
    \centering
    \caption{Classification accuracy of \sol\ considering 15 SanDisk 16GB USB Flash Drives.}
    \label{fig:scenario2_sandisk}
\end{figure}

\begin{figure}[htbp]
    \includegraphics[width=.6\columnwidth]{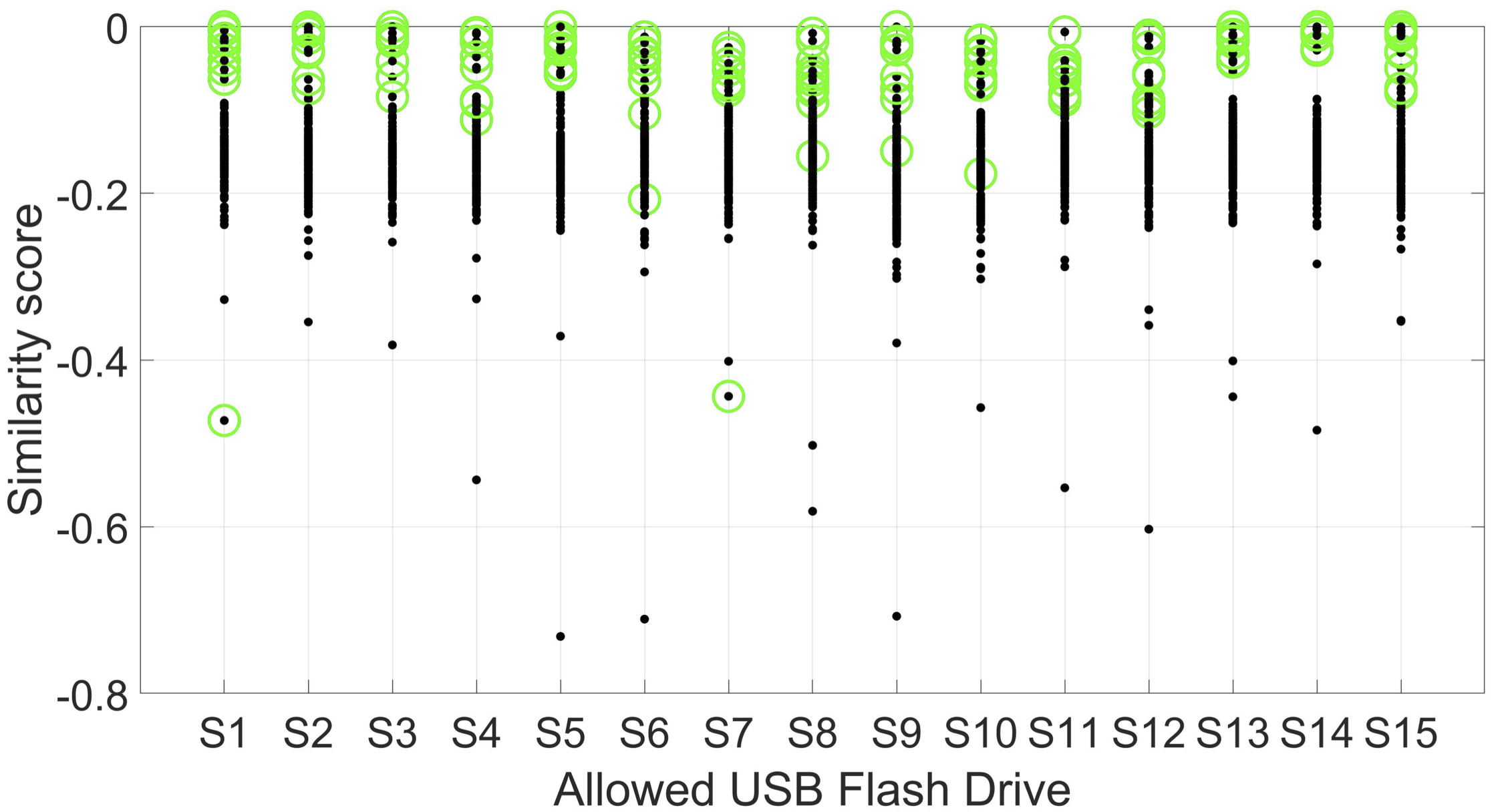}
    \centering
    \caption{Classification accuracy of \sol\ considering 15 Strontium 16GB USB Flash Drives.}
    \label{fig:scenario2_strontium}
\end{figure}

\begin{figure}[htbp]
    \includegraphics[width=.6\columnwidth]{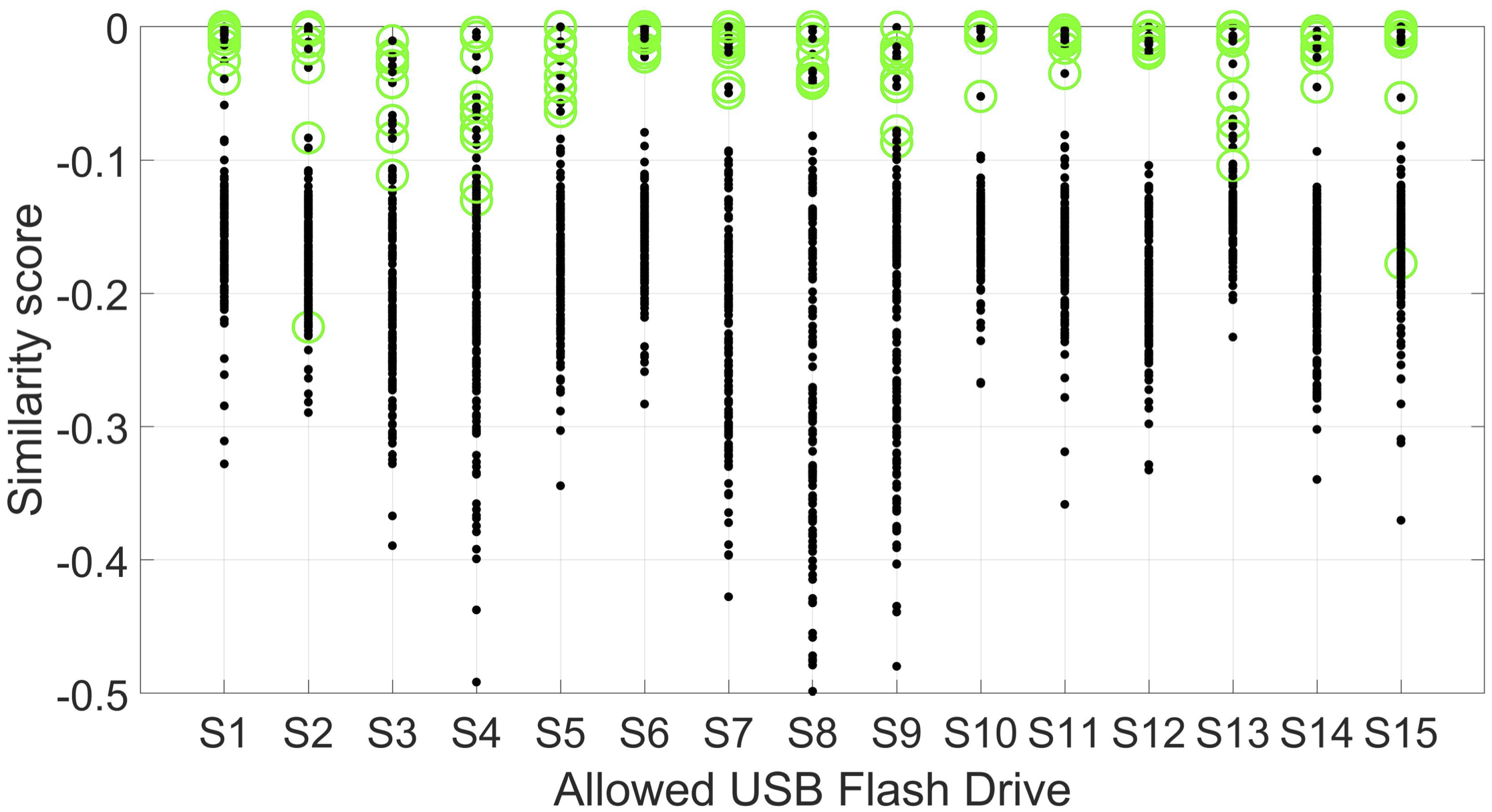}
    \centering
    \caption{Classification accuracy of \sol\ considering 15 Klevv Neo C20 16GB USB Flash Drives.}
    \label{fig:scenario2_klevv}
\end{figure}

We considered three different USB Flash Drive brands: the SanDisk 16GB (U15), the Strontium 16GB (U16), and the Klevv Neo C20 16GB (U17), where $15$ USB flash drives were available for each experimentation. The results are reported in Figures \ref{fig:scenario2_sandisk},~\ref{fig:scenario2_strontium} and \ref{fig:scenario2_klevv} for the SanDisk, the Strontium and the Klevv Neo C20, respectively. Note that, now, each column contains $15 \cdot 10 = 150$ experiments, where $10$ experiments refer to the authorized USB flash drive and $140$ experiments refer to the unauthorized drives. \\
Figures~\ref{fig:scenario2_sandisk}-\ref{fig:scenario2_klevv} show that most of the (authorized) green circles are characterized by evaluation scores higher than the black (unauthorized) dots. Generally speaking, the selection of the threshold is a key component to determine the accuracy of the devised brands classification method. Low values of the threshold guarantee zero false positives, i.e., no unauthorized brands are erroneously recognized as authorized, but it also implies an increased number of false negatives, i.e., authorized brands erroneously recognized as unauthorized, and thus experiencing a reduced number of True Positives (correctly authorized brands). As a worst-case, we notice that selecting a common threshold $T=-0.09$ for all the SanDisk, the Strontium, and the Klevv Neo, it is possible to guarantee a True Positive Ratio of $0.946$, $0.913$, and $0.963$, and a False Positive Ratio of $0.019$, $0.027$, and $0.012$, respectively. We notice that this is the worst-case, as a threshold can be selected appropriately for each single USB Flash Drive. As an example, with the Strontium USB flash drive, an appropriate choice of the thresholds for each flash drive (that can be done looking at Figure~\ref{fig:scenario2_strontium}) can provide up to TPR $0.94$ and FPR $0.005$. Overall, based on the specific USB flash drive, the network administrator can setup \sol\ in a way to achieve the best trade-off between the TPR and the FPR, to minimize the false positives. In addition, the administrator can also suggest a specific brand to be used for authentication, based on the achievable ratio between the TPR and the FPR.

Finally, we remark that discriminating against the specific USB device requires more powerful equipment than a simple SDR. Specifically, the equipment to be used should be characterized by a wide real-time resolution bandwidth, of about 200 MHz. Despite in our experimental assessment we used the \emph{Rohde \& Schwarz FSW8} spectrum analyzer, other cheaper devices are available, characterized by a similar bandwidth range, with a price starting from as low as 500 USD.

\subsection{Considerations on Features Reduction}
\label{sec:exp_fr}
As previously highlighted, \sol\ requires a total number of 325 features to achieve the reported performances. The reported number of features could appear high, with the risk of incurring in the curse of dimensionality and subsequent model overfitting.

First, we highlight that the methodology adopted by \sol\ is based on the split of the whole data set into multiple disjoint training and test sets. Specifically, for each test, the model of each USB Flash Drive is trained on $9$ samples and tested on the remaining one. This approach is then iterated, by testing each available sample as a test set (while the others serve as a training set). It is worth noting that the model has not been updated after the testing, and we did not refine the hyper-parameters of the one-class SVM classifier (being these two among the common sources of overfitting). Therefore, we can reasonably exclude the hypothesis of overfitting.

However, to provide further insights, we investigated the effectiveness of \sol\ using a reduced number of features, both for Scenario 1 and a reference example for Scenario 2. To determine the critical features, we rank them with the help of the FEAST toolbox, which is a commonly used feature ranking tool in machine learning~\cite{brown2012_jmlr}.  This toolbox has been used also recently by the authors in~\cite{Cheng2019_ccs}, to reduce the dimensionality of the features in the context of unintentional magnetic emissions.
The following Table~\ref{tab:scenario1_fr} and Table~\ref{tab:scenario2_fr} show the results of our investigation with a different number of pre-dominant features, assuming the Scenario 1 (brand and model identification) and the worst-case of Scenario 2 (USB authentication, 15 Strontium USB Flash Drives), respectively. In the tables, we also report the correspondent selection of the threshold value (worst-case), set up to achieve the reported results.

\begin{table}[htbp]
\caption{True Positive Ratio (TPR) and False Positive Ratio (FPR) of \sol\ for the Scenario 1 (brand and model identification), using an increasing number of pre-dominant features, obtained with the FEAST toolbox.
}
\centering
\color{black}
    \begin{tabular}{|c|P{3.0cm}|P{2.2cm}|P{3.1cm}|}
    \hline
        \textbf{No. of Features} & \textbf{Threshold Value} & \textbf{TPR} & \textbf{FPR} \\ \hline
        10     & T = -0.015 & 0.770 & 0.0003 \\
        20     & T = -0.015 & 0.935 & 0.0014 \\
        50     & T = -0.06  & 0.929 & 0.0018 \\
        100    & T = -0.07  & 0.911 & 0.0025 \\
        200    & T = -0.07  & 0.947 & 0.0029 \\
        300    & T = -0.078 & 0.976 & 0.0051 \\
        325    & T = -0.078 & 0.982 & 0.0022 \\
        \hline
    \end{tabular}
    \label{tab:scenario1_fr}
\end{table}
\begin{table}[htbp]
\caption{True Positive Ratio (TPR) and False Positive Ratio (FPR) of \sol\ for the Scenario 2 (USB authentication), in the case of 15 Strontium USB flash drives, using an increasing number of pre-dominant features, obtained with the FEAST toolbox.
}
\centering
\color{black}
    \begin{tabular}{|c|P{3.0cm}|P{2.2cm}|P{3.1cm}|}
    \hline
        \textbf{No. of Features} & \textbf{Threshold Value} & \textbf{TPR} & \textbf{FPR} \\ \hline
        10     & T = -0.8  & 0.753 & 0.040 \\
        20     & T = -0.9  & 0.78  & 0.026 \\
        50     & T = -0.9  & 0.78  & 0.026 \\
        100    & T = -0.9  & 0.893 & 0.041 \\
        200    & T = -0.93 & 0.893 & 0.028 \\
        300    & T = -0.93 & 0.913 & 0.027 \\
        325    & T = -0.93 & 0.913 & 0.027 \\
        \hline
    \end{tabular}
    \label{tab:scenario2_fr}
\end{table}
We notice that the fluctuations of the results by increasing the number of features is limited. For the Scenario 1, the most representative 20 features are enough to obtain a TPR of $0.93$, with a negligible fraction of false positive (FPR $0.0014$). At the same time, for the case of Scenario 2, assuming to work with the Strontium USB Flash Drives, $100$ features are enough to obtain a TPR of $0.893$, and including more features has only a slight impact on the performance accuracy.

Based on the above considerations and results, we believe that we are both avoiding the curse of dimensionality, and not overfitting the model.

\subsection{Discussion on Firmware Modifications}
\label{sec:fw_mod}
As highlighted in Section~\ref{sec:scenario_adv_model}, the firmware of commercial USB flash drives is usually protected by multiple security layers, and cannot be modified. Despite several tutorials are available on media and technical blogs regarding firmware modification of commercial USB flash drives, this is possible only due to vulnerabilities affecting specific versions of the firmware of the USB flash drive. As a direct consequence, the aforementioned vulnerabilities are usually (and immediately) fixed by the vendor.

However, there are specific USB flash drives, such as the Rubber Ducky, whose firmware is specifically designed to be modified to achieve several tasks. These tasks can include benign operations, such as the opening of a file, or malicious activities, such as the disabling of anti-virus software on the host, or the activation of passwords stealing tool.

\begin{figure}[htbp]
    \includegraphics[width=.6\columnwidth]{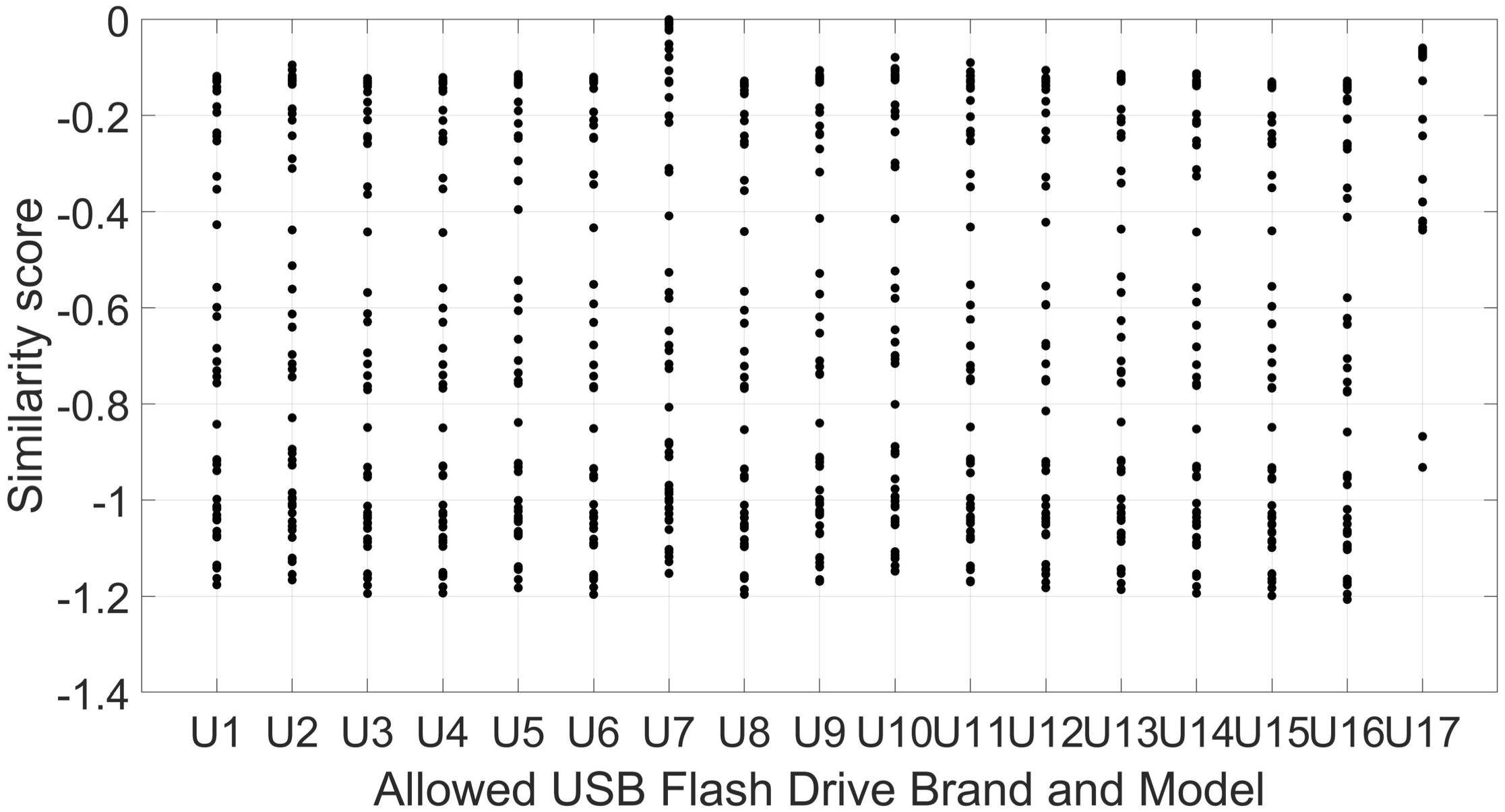}
    \centering
    \caption{Classification accuracy of \sol\ considering 6 modified versions of the boot firmware of the Rubber Ducky USB Flash Drive (U7).}
    \label{fig:modifiedRD_all}
\end{figure}

To provide more insights on the performance of \sol\ to identify modified firmware versions, we investigated the effectiveness of our solution considering seven (7) different firmware versions of the Rubber Ducky (U7):
\begin{itemize}
    \item F1. Baseline default firmware used in Section~\ref{sec:class_usb}, injecting an empty payload; 
    \item F2. Firmware that opens a text file and types a single \emph{Hello World} line;
    \item F3. Firmware that opens a text file and types 100 \emph{Hello World} lines;
    \item F4. Firmware that opens a text file and types 200 \emph{Hello World} lines;
    \item F5. Firmware that disables Windows Defender;
    \item F6. Firmware that inserts a delay of approximately $750$~ms;
    \item F7. Firmware that steals browser-stored passwords.
\end{itemize}
First, we investigated the performance of \sol\ to identify the above firmware modifications, without any previous knowledge of their existence. To this aim, we used the models created for the tests in Section~\ref{sec:class_usb}, while the test set consisted of $10$ acquisitions for the firmware versions \{$F2, \ldots, F7$\}. The results are reported in Figure~\ref{fig:modifiedRD_all}. Note that all the markers are now black dots, as the whole test set samples are considered to be \emph{unauthorized}.
First, we notice that for all the experiments that assume as \emph{authorized} a USB flash drive that is different than the \emph{U7}, report values that are lower than $-0.01$. 
Considering the configuration of the threshold discussed in Section~\ref{sec:class_usb}, these experiments all provide correct \emph{True Negative} outputs. Considering the USB flash drive $U7$, we notice that, out of $50$ experiments, $13$ report values that exceed the local threshold $T=-0.0365$, i.e., the local threshold value obtained from the experiments in Section~\ref{sec:class_usb}, providing FPR $0$. 
These experiments correspond to the firmware $F2$ (8 experiments) and $F5$ (5 experiments). Therefore, in the absence of any training on the modified firmware versions, \sol\ can discriminate a modified firmware only if the profile of the corresponding EM emanations is consistently different from the \emph{trained} one. When the modifications are very little, such as for $F2$ and $F5$, the accuracy of \sol\ decreases.

\begin{figure}[htbp]
    \includegraphics[width=.6\columnwidth]{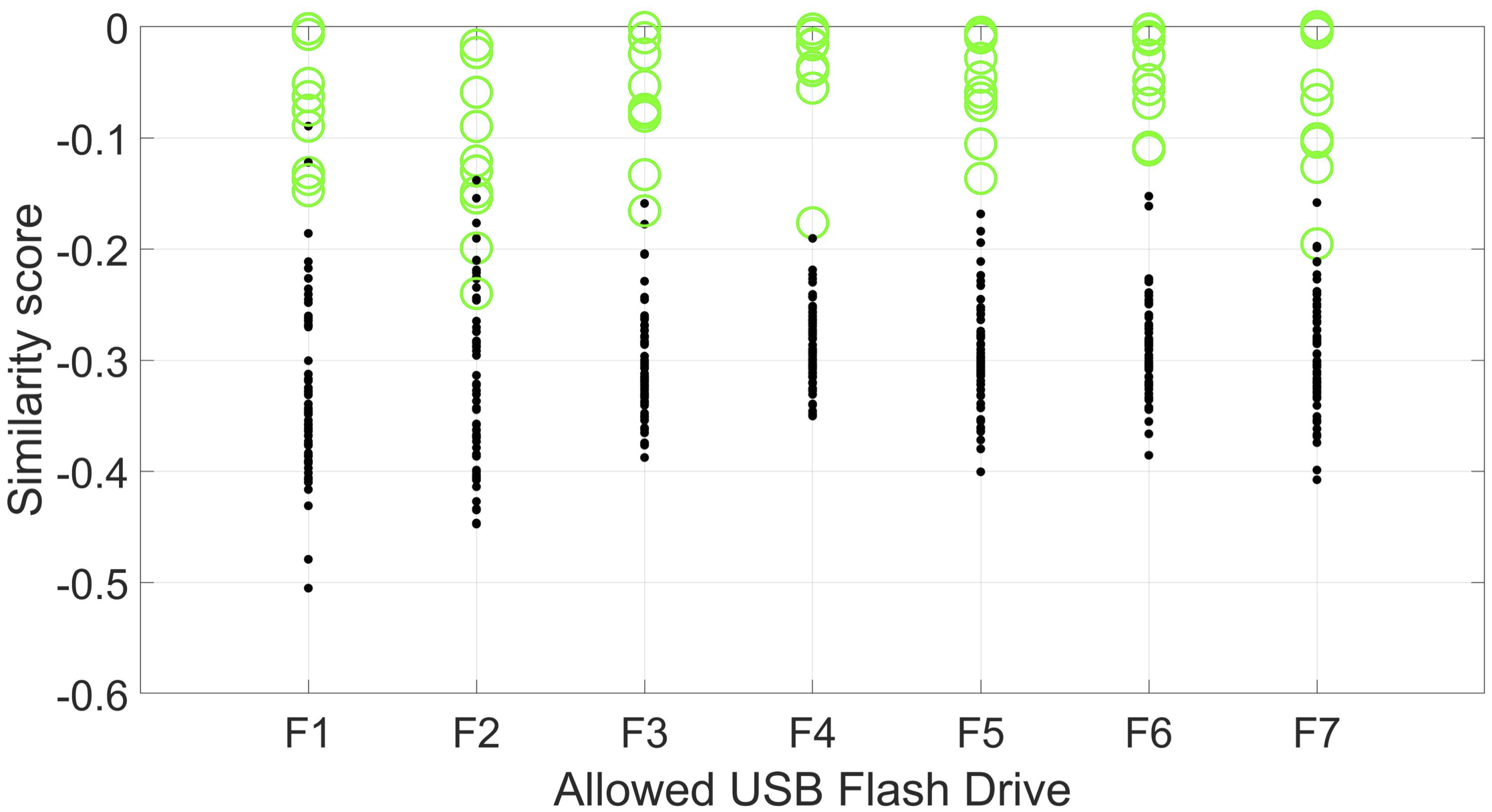}
    \centering
    \caption{Classification accuracy of \sol\ considering 6 modified versions of the boot firmware of the Rubber Ducky USB Flash Drive (U7), assuming to train on the modified firmware versions.}
    \label{fig:modifiedRD_trained}
\end{figure}

However, we notice that the accuracy of \sol\ can be significantly enhanced by training on the modified firmware versions. When this is possible (as for the Rubber Ducky), Figure~\ref{fig:modifiedRD_trained} provides the results of such a scenario, using the same procedure used for the previous sections.
It is worth noting that selecting the previously adopted threshold value, i.e. $T=-0.07$, \sol\ achieves FPR $0$, i.e., it does not mislead any of the firmware versions with the native firmware version $F1$ for the USB flash drive $U7$.

\subsection{Host Device (Verifier) configuration and its impact on \sol.}
\label{sec:host_impact}

The analysis and results reported in the previous sections have been obtained using a single host device, as detailed in Section~\ref{subsec:tools}.
This is consistent with the two scenarios and the system model assumed in our work. Indeed, both in a company (Scenario \#1) and in a critical infrastructure (Scenario \#2), it is reasonable to assume that the system administrator has full control of the verifier, i.e., the system used to acquire the fingerprints of the legitimate USB devices and to test new USB devices. Moreover, it is reasonable and convenient for the system administrator that such a system uses the same \acl{OS} of the system(s) used within the site.

To provide further insights about the fingerprinting process, in this subsection, we extend our analysis, investigating the consistency of the fingerprinting process when the hardware and the software of the host devices change.

\begin{figure}[htbp]
\centering
    \includegraphics[width=.45\columnwidth]{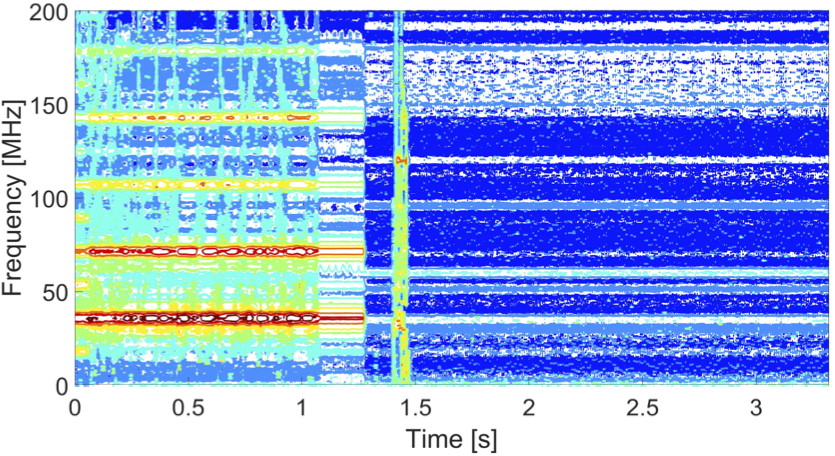}
    \includegraphics[width=.45\columnwidth]{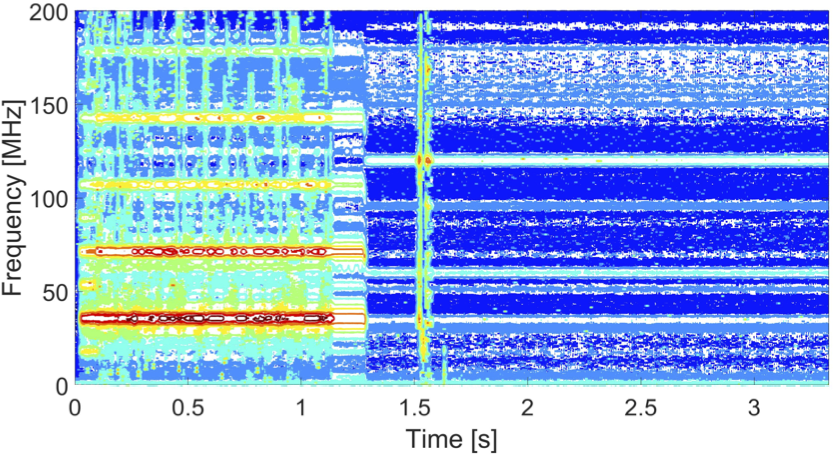}
    \centering
    \caption{Magnetic emissions of a USB flash drive (SanDisk Cruzer 128GB, U10), on (a) Windows 10 OS; and, (b) Windows 7 OS.}
    \label{fig:diff_os}
\end{figure}

As a reference example, we investigated the consistency of the magnetic emissions radiated during the boot process by a USB flash drive when different versions of the same OS are used on the same machine. To this aim, we collected the unintentional magnetic emissions radiated during the boot of a sample USB flash drive, i.e., the SanDisk Cruzer 128GB drive (U10), when connected to the same machine used for the previous tests, operating with the Windows 7 and the Windows 10 OS. The former includes the version 6.1.7601.1910 of the USB driver, dating back to June 2006, while the latter mounts the latest available version at the time of this writing, i.e., the version 10.0.18362.1 of the USB driver (March 2019). The profiles are shown in Fig.~\ref{fig:diff_os}.

Comparing the two figures, some small differences are noticeable, especially in the duration of the different phases of the boot sequence.
We remark that, while we report the figures only for the $U10$ flash drive, the above-described behavior is common to all the USB flash drives analyzed in our work.

Overall, these findings suggest that the construction of the Local Database (i.e., the training process) and the Classification (i.e., the testing process) required by \sol\ should be carried out on the same system, guaranteeing both hardware and software consistency. When at least one of these features changes, differences arise that, especially in the case of Scenario \# 2, could lead to undesired mismatches. Conversely, when the system administrator only requires the verification of the brand and the model of the USB flash drive (Scenario \# 1), fingerprints can be acquired and tested by using different systems or different OSs.

We highlight that these findings are consistent with other related work on unintentional magnetic emissions. Indeed, small modifications to the hardware and software features of involved devices lead to different fingerprints.

Overall, our analysis demonstrates that the profile of the unintentional magnetic emissions radiated during the boot procedure is mainly caused by the combination of four main elements: (i) the memory readings/writings of the USB flash drive; (ii) the software instructions executed by the microcontroller chip of the USB flash drive; (iii) the layout of the PCB of the USB flash drive; and, (iv) the hardware and software features of the host device. 
This is confirmed by several findings. Considering the same micro-controller chip on a specific host, we have recorded completely different profiles of unintentional magnetic emissions when considering different USB flash drives, characterized by different layout of the PCB. This can be verified by looking at the profile of the unintentional emissions of the devices HP (U1), Kingston (U3), PNY (U5), Patriot (U6), Silicon Power (U13), and Toshiba (U14), sharing the same micro-controller chip but a different layout of the PCB.

In addition, as demonstrated by the above analysis, changing the hardware and the software of the host device has a small yet noticeable impact on the fingerprints. When only the software of the host changes, differences in the fingerprint of the same USB flash drive become more pronounced due to a different interaction of the OS with the USB flash drive, affecting the final part of the boot procedure. Considering different hardware (USB flash drive) on the same physical machine, we also experienced small differences in the final part of the boot. 

The discussion reported above highlights the effect of the hardware and the software of the host device on the overall recorded fingerprint. However, we remark that the peculiarities associated with the magnetic emissions are both out of the scope of our work and technically hard to achieve when working with System-on-Chip (SoC), such as the USB flash drives.
Indeed, matching the effect of every single electronic component to the fingerprint would require isolating the effect of each of them on the PCB of the specific USB flash drive. This is very hard to achieve both via hardware and software. As for the hardware, disconnecting its components turns out to be very difficult to achieve, in particular when working with SoC, such as the USB flash drives, where all the electronic components are integrated on the same PCB. Disabling the components via software, instead, would require having strict control over the firmware and the software of the USB flash drive. However, they are not accessible by end-users and protected against source code reverse-engineering by multiple security layers.

Overall, we can conclude that the differences among the brands and the devices of the same brand are due to small circuits differences, as well as inaccuracies and imperfections in the manufacturing process. On the one hand, the underlying process behind MAGNETO can be brought back to a specific use-case of the Electro-Magnetic emanations of embedded circuits. On the other hand, we would like to recall that our work aims to identify either the brand and the model or the specific USB device, including all the interacting components, as a unique device. Thus, analyzing the effect of each component of the leakage is out of scope for our work. We also remark that \sol\ is novel compared to other related work as, to the best of the authors' knowledge, it is the first solution that can fingerprint USB flash drives in a non-interactive, minimal-invasive, and privacy-preserving fashion, not requiring any intervention or modification on the devices under test.\\
We also highlight that \sol\ can be further extended to interact with the host OS, to instruct the sampling of the unintended magnetic emissions radiated by a connected USB flash drive asynchronously, e.g., when a suspected activity is recorded by the host OS via software. In such a scenario, it could be possible to compare the recorded profile of the specific USB flash drive in regular operating conditions with the actual one, to identify eventual deviations. However, this extension is left as future work.
Finally, we highlight that our methodology, i.e., fingerprinting the device as a whole, without considering the effect of the single components, is consistent with all the related work investigating physical-layer fingerprinting and unintentional Electro-Magnetic emissions, including ~\cite{Dejean2007, camurati2018, cobb2010, Cobb2012_tifs, wright2014, bihl2016, Dubendorfer2012, ramsey2012, Suski2008, lukacs2015, Cheng2019_ccs}. The physical effects underpinning these phenomena are described by the physics fundamental Maxwell equations, and also specific mathematical models are available in the literature, describing their complex creation \cite{bole2009}.

\section{Discussion and Limitations}
\label{sec:discussion}

The results presented in Section \ref{sec:experiments} clearly showed the effectiveness of \sol. We showed that, considering a specific host device, the unintentional magnetic emissions radiated at low frequencies during the boot by a USB flash drive are: (i) very similar for devices of the same brands, (ii) unique for each device, and (iii) consistent in time, thus being a viable and effective tool for USB brand and device fingerprinting. 
For instance, assuming the reference scenario of a company, such as the \emph{Scenario \# 1} discussed in Section \ref{sec:scenario_adv_model}, the system administrator can enforce the use of only specific brands and models of USB flash drive, eventually provided by the company itself. To this aim, the system administrator can deploy \sol\ on a dedicated machine, acquiring the profile of the unintentional magnetic emissions of the selected USB flash drives, before their use.
Then, in the case of a company, on a selected period (e.g., once a day), every time a USB flash drive needs to be used, the system administrator can plug the device into a dedicated laptop, connected to the proposed \sol\ system. Then, if the magnetic emissions at low frequencies during the boot operations match the stored profile for the particular brand and model, the use of the particular USB flash drive could be authorized; otherwise, it will not. 
Similarly, in a critical infrastructure, such as in the \emph{Scenario \# 2} discussed in Section \ref{sec:scenario_adv_model}, it is possible to deploy \sol\ by using powerful equipment, able to acquire the profile of the magnetic emissions on a wider frequency span. As shown in Section \ref{sec:auth}, using such an increased frequency span can enhance the capabilities of \sol, to provide the identification of the specific USB drive.
\\
\noindent
Another feature of \sol\ is its \emph{linear overhead}, independent of the amount of USB flash drives on the market. 
Indeed, the system administrator needs only to 
generate the profile (according to the procedure detailed in the paper) of the USB he/she wants to add to the whitelist of the authorized USB drives. Once the process committed, any USB drive that is screened requires just the collection of the drive's magnetic profile and its comparison against the profiles in the white-list. The operation has a cost that is linear in the size of the white-list.

\noindent
It is worth noting that the setup of \sol\ is only aimed at protecting and ensuring the authenticity of the boot procedures. This choice allows to timely detect a malicious adversary that: (i) replaces the regular USB flash drive with another one having the same external look and form factor, but a different hardware and (possibly malicious) firmware; and, (ii) can replace the firmware of a legacy USB device with a brand new one, by inserting its own code. Such strategies are common to the vast majority of USB attacks known today, as they are based on the complete replacement of the hardware or the firmware of a regular USB device \cite{Nissim2017}.
\\
{\bf USB firmware tampering.} Despite some USB flash drives are certified according to FIPS 140-2 Level 2 and Level 3 standards, USB flash drives cannot be considered as protected against tampering. As demonstrated in Section~\ref{sec:fw_mod}, the strength of \sol\ lies in the inherent capability to detect the tampering, as long as the modified firmware causes a profile of the unintended magnetic emissions sufficiently different from the recorded one.\\
Indeed, the adversary might replace the firmware of the USB flash drive with a malicious one, where just the minimum amount of lines have been modified to trigger a specific malware. Despite \sol\ can deal with these modifications by training on the modified firmware versions, we highlight that both the executable file and the source code of the firmware on-board of commercial USB flash drives are usually not modifiable and not accessible by the end-users after the deployment. Even in the unlikely case where an adversary can obtain the file of the compiled firmware, it is hard and impractical to reverse-engineer it to obtain the source code. Indeed, being always protected by intellectual property rights, the source code of such firmware is also secret, protected by multiple security layers deployed at manufacturing time, and not available for public download. Thus, it is unfeasible to obtain it and to deploy the aforementioned attack.

\noindent
Similarly, another adversarial strategy could consist in smartly modifying the firmware of the legacy USB device, inserting just a code line that postpones the injection of the malicious code after the execution of the boot, i.e., when the USB drive goes in the idle state, waiting for instructions by the host. 
Despite the considerations previously done for the legacy firmware modifications are still applicable to this attack, we also observe that \sol\ can be extended to detect the previously mentioned attack. By simply widening the observation window of the \emph{Emissions Extraction Module} and recording the unintentional magnetic emissions when the USB device is in the idle state, \sol\ can detect any anomalous operations initiated by the USB device, raising an immediate alarm. Thus, \sol\ can be extended to guarantee the safe usage of any USB flash drive over an arbitrarily large time after the USB device has been connected to the host system. 

\noindent
More powerful adversaries could adopt more sophisticated strategies. For instance, they could tie the execution of the malicious code to a file transfer operation on the particular USB flash drive. When a file is transferred to (or from) the drive, the malicious code is triggered. The effectiveness of \sol\ against this attack could still be achieved by fingerprinting also a file transfer operation, assuming that a file transfer of a given size leads to the same profile of the unintended radiated emission. To neutralize our countermeasure, the adversary could always trigger the execution of the malicious code in a way to be tied to the size of the file to be transferred (e.g., greater than 10 MB). In this case, \sol\ would be effective only if the profile of the unintentional magnetic emissions for the specific file transfer operation has been recorded and profiled before, during the \emph{Training Mode}. It follows that fingerprinting all these operations would affect also the latency and the usability of \sol, increasing the testing time and the overall size of \sol. 
However, if the attack is triggered only when a specific file stored on the USB is opened, \sol\ cannot be effective without any previous fingerprinting process on that specific file.

\section{Conclusion}
\label{sec:conclusion}

In this paper, we proposed \sol, a framework able to enhance the security level in the usage of USB flash drives. Depending on the used equipment, \sol\ can uniquely identify either the brand and the model of commercial USB flash drives, or the specific USB flash drive, by analyzing unintentional magnetic emissions radiated by the target USB devices during the execution of the boot procedure. 
Through extensive experimental measurements on 59 different USB drives---belonging to 17 brands, including the top brands purchased on Amazon in mid-2019--- we demonstrated that a USB flash drive can be fingerprinted by only looking at the magnetic emissions unintentionally radiated during the boot procedure on a given host. 
When coupled with a commercial low-cost \acl{SDR} such as the HackRFOne, \sol\ can identify the specific brand and model of the USB flash drive with a minimum classification accuracy of the $98.2$\%, guaranteeing, at the same time, only $0.01$\% of false positives. When coupled with more expensive equipment, characterized by a wide analysis bandwidth of at least 200 MHz, \sol\ can also identify the specific USB flash drive with a minimum classification accuracy of $91.3$\%. All the results above can be obtained in almost real-time, analyzing a time-frame of at most $3.35$ seconds and requiring a negligible processing time (always less than 1 second on a standard laptop).

Thanks to the reported outstanding performance, \sol\ emerges as a viable option to strengthen the security of any sensitive computing equipment exposed to the threats posed by USB drives, such as the one in a telco-backbone or a power plant, and in general contributing to the security of critical infrastructure systems.

\section*{acknowledgments}
The authors would like to thank the anonymous reviewers, that helped to improve the quality of the manuscript.
This publication was made possible by awards NPRP11S-0109-180242 and GSRA6-1-0528-19046, from the QNRF-Qatar National Research Fund, a member of Qatar Foundation. The findings achieved herein are solely the responsibility of the authors.

\bibliographystyle{ACM-Reference-Format}
\bibliography{bibliography}

\end{document}